\documentclass{aa} 
\usepackage{graphicx}
\usepackage{txfonts}
\usepackage{float}
\usepackage{amsmath, amssymb}
\usepackage{epstopdf}
\usepackage{longtable}
\usepackage{array}
\usepackage{color}

\usepackage{threeparttable}
\bibpunct{(}{)}{;}{a}{}{,} 
\newcommand{\Teff}{\ensuremath{T_{\mathrm{eff}}}}
\newcommand{\logg}{\ensuremath{\log g}}
\newcommand{\feh}{\ensuremath{[\mathrm{Fe/H]}}}
\newcommand{\afe}{\ensuremath{[\mathrm{\alpha/Fe]}}}

\newcommand{\lrlhk}{\ensuremath{\log(R^\prime_{\rm HK})}}
\newcommand{\lrlhkt}{\ensuremath{\log R^\prime_{\rm HK} (T_{\rm eff})}}
\newcommand{\lrlhkb}{\ensuremath{\log R^\prime_{\rm HK} ({\rm B-V})}}

\newcommand{\smw}    {\ensuremath{S_{\mathrm MW}}}
\newcommand{\vsini}{\ensuremath{v \sin i}}
\newcommand{\Ha}{H$\alpha$}
\newcommand{\Hb}{H$\beta$}

\newcommand{\HK}{\ion{Ca}{II}~} 

\newcommand{\hk}{\ion{Mg}{II}~}
\begin{document}

   \title{The Solar Twin Planet Search: The age - chromospheric activity relation \thanks{Based on observations collected at the European Organisation
for Astronomical Research in the Southern Hemisphere under ESO
programs 188.C-0265, 183.D-0729, 292.C-5004, 097.C-0571, 092.C-0721, 093.C-0409, 072.C-0488, 183.C-0972, 091.C-0936, 192.C-0852, 196.C-1006, 076.C-0155, 096.C-0499, 185.D-0056, 192.C-0224, 075.C-0332, 090.C-0421, 091.C-0034, 077.C-0364, 089.C-0415, 60.A-9036, 092.C-0832, 295.C-5035, 295.C-5031, 60.A-9700, 289.D-5015, 096.C-0210, 086.C-0284, 088.C-0323, 0100.D-0444, and 099.C-0491.}}

   \author{Diego Lorenzo-Oliveira\inst{1}
	  \and
	  Fabrício C. Freitas\inst{1}
          \and
          Jorge Mel\'{e}ndez\inst{1}
           \and
          Megan Bedell\inst{2,7}
          \and
	Iv\'{a}n Ram{\'{\i}}rez\inst{3}
          \and
          Jacob L. Bean\inst{2}
          \and
          Martin Asplund\inst{4}
          \and
          Lorenzo Spina\inst{1,8}
	\and
          Stefan Dreizler\inst{5}
	\and
          Alan Alves-Brito\inst{6}
	\and
          Luca Casagrande\inst{4}
          }

   \institute{Universidade de S\~ao Paulo, Departamento de Astronomia do IAG/USP, Rua do Mat\~ao 1226, 
              Cidade Universit\'aria, 05508-900 S\~ao Paulo, SP, Brazil. \email{diegolorenzo@usp.br}
         \and
             University of Chicago, Department of Astronomy and Astrophysics, USA
         \and
             Tacoma Community College, 6501 South 19th Street, Tacoma, Washington 98466, USA
         \and
             The Australian National University, Research School of Astronomy and Astrophysics, Cotter Road, Weston, ACT 2611, Australia
         \and
            Institut f\"{u}r Astrophysik, Universit\"{a}t G\"{o}ttingen, Germany
         \and
            Instituto de Fisica, Universidade Federal do Rio Grande do Sul, Porto Alegre, Brazil
            \and
            Center for Computational Astrophysics, Flatiron Institute, 162 5th Ave., New York, NY 10010, USA 
            \and
            Monash Centre for Astrophysics, School of Physics and Astronomy, Monash University, VIC 3800, Australia}
   
  \abstract
   {It is well known that the magnetic activity of solar type stars decreases with age, but it is widely debated in the literature whether there is a smooth decline or if there is an early sharp drop until 1-2 Gyr followed by a relatively inactive constant phase.}
   {We revisited the activity-age relation using time-series observations of a large sample of solar twins whose precise isochronal ages and other important physical parameters have been determined.}
   {We measured the \HK H and K activity indices using $\approx$ 9000 \rm{HARPS} spectra of 82 solar twins. In addition, the average solar activity was calculated through asteroids and Moon reflection spectra using the same instrumentation. Thus, we transformed our activity indices into the S Mount Wilson scale (\smw), recalibrated the MW absolute flux and photospheric correction equations as a function of \Teff, and then computed an improved bolometric flux normalized activity index \lrlhkt\,for the entire sample.}
   {New relations between activity and age of solar twins were derived assessing the chromospheric age-dating limits using \lrlhkt. We measured an average solar activity of \smw = 0.1712 $\pm$ 0.0017 during solar magnetic cycles 23$-$24 covered by HARPS observations and we also inferred an average of \smw = 0.1694 $\pm$ 0.0025 for cycles 10$-$24, anchored on a S index vs. sunspot number correlation. Also, a simple relation between the average and dispersion of the activity levels of solar twins was found. This enabled us to predict the stellar variability effects on the age-activity diagram and, consequently, estimate the chromospheric age uncertainties due to the same phenomena. The age-activity relation is still statistically significant up to ages around 6$-$7 Gyr, in agreement with previous works using open clusters and field stars with precise ages.}
   {Our research confirms that \HK H\& K lines remain a useful chromospheric evolution tracer until stars reach ages of at least 6$-$7 Gyr. We found an evidence that, for the most homogenous set of old stars, the chromospheric activity indices seem to continue decreasing after the solar age towards the end of the main-sequence. Our results indicate that a significant part of the scatter observed in the age-activity relation of solar twins can be attributed to stellar cycle modulations effects. The Sun seems to have a normal activity level and variability for its age.}

   \maketitle
%

\section{Introduction}

The solar-type stars' chromospheric activity is one of the observed manifestations of a broad phenomenon called stellar magnetic activity, which is expected to be driven by the same physical principles of the Solar dynamo. The paradigm is that the complex interplay between turbulent convection and rotation triggers the stellar cyclic and self-sustained global magnetic activity \citep{parker70}. As the star ages, it is expected that its rotation and, consequently, magnetic activity, decreases due to angular momentum loss through magnetized winds and structural variations along evolutionary timescales. 

Therefore, considering this theoretical framework, rotation \citep{skumanich72, barnes07, barnes10, reiners12, santos16} and magnetic activity \citep{skumanich72, soderblom91, mamajek08, lorenzo16} are frequently considered as interesting clocks optimized for main-sequence solar-type mass stars. Alternatively, some authors estimate stellar ages using classical tecniques such as isochrones \citep[e.g.,][]{ng98,lachaume99,ramirez14,nissen15}, and the use of chemical abundance markers such as the Li abundance \citep[e.g.,][]{skumanich72,donascimento09,carlos16} or, more recently, the [Y/Mg] or [Y/Al] ratio \citep[e.g.,][]{nissen15,tuccimaia16,spina16a,spina16b}. A review of different methods to estimate stellar ages is given by \citet{soderblom10}, who also discussed the problems affecting the different age indicators. 
   
The first parametrization of the activity-age relation was performed by \citet{skumanich72}, where the chromospheric emission of the \HK H \& K lines was used as activity indicator. While there are other important magnetic activity tracers such as high-energy coronal emissions \citep{ribas05, booth17}, \hk h \& k \citep{oranje85, buccino08}, \Ha\, \citep{pasquini91, lyra05}, \Hb\, \citep{montes01}, \HK infrared triplet \citep{busa07,lorenzo16b}, the \HK H \& K lines are widely used because they are readily measurable from ground-base observatories, and also there is a consistent and ready-to-use absolute flux calibration available in the literature.

Most of the previous works suggested a smooth decrease of chromospheric activity with increasing age \citep{soderblom91,mamajek08,lorenzo16}, but \citet{pace04} and \citet{pace13} suggested that activity-age relations are only valid for stars younger than approximately 1.5 Gyr, with no further decay in activity after this age. The authors analysed high-resolution {\rm UVES} observations of 35 FG-type members of 5 open clusters spanning a wide age interval from Hyades to M67, and the Sun, resulting in a fit between chromospheric flux or \vsini, and age. In addition, \citet{pace13} assessed the age-activity diagram also through hundreds of FGK dwarfs with photometric effective temperatures and metallicities from \citet{casagrande11}, and open clusters, to indicate a plateau after $\sim$ 1.5 Gyr. This result, combining a heterogeneous sample of field stars and open clusters is an independent confirmation of previous findings of \citet{lyra05} using the \Ha\,line, and is also in line with \citet{pace04}. 

On the other hand, recent work by \cite{lorenzo16} using dozens of M67 ($\approx$ 4 Gyr) and NGC 188 ($\approx$ 6 Gyr) G dwarfs observed with {\rm Gemini North GMOS} shows that the activity evolution could be extended until at least 6 Gyr. Furthermore, the authors point out that the lack of activity evolution after 1.5 Gyr could be interpretated as a mass/effective temperature and metallicity bias that affects \HK H \& K fluxes, in addition to isochronal age sample selection bias.

Therefore, the most convenient way to rule out the effects of other variables on \HK\,activity levels \citep{rutten87,rochapinto98, gray06, lovis11, lorenzo16b}, minimizing biases in the age-activity correlation, is the study of chromospheric activity evolution of open clusters members \citep{soderblom91,mamajek08}, wide binaries \citep{garces11,desidera06}, or field stellar twins with similar mass and metallicity. In this work we adopted the last option, reassessing the age-chromospheric activity relation using a large sample of solar twins \citep{ramirez14}. These stars are very similar to the Sun, since their stellar parameters (\Teff, \logg, \feh) are roughly within $\pm$100 K, $\pm$0.1 dex, $\pm$0.1 dex of the Sun's values \footnote{\Teff$^\odot$ = 5777K, \logg$^\odot$ = 4.437, as adopted in \citet{ramirez14}.}. As the stars have very similar physical properties (mass and metallicity), the main parameter affecting changes in stellar activity is their ages. In a broader context, the magnetic activity history of our Sun is important for planetary habitability \citep{ribas05,airapetian16,donascimento16} and to constrain dynamo models \citep[e.g.,][]{karak14,pipin16}.

This paper is organised as follows: Sec. \ref{sec:data} describes our working sample and the procedures adopted to build a new \HK H \& K chromospheric activity index. We also investigate the solar activity variability in comparison to the solar twins. In Sec. \ref{sec:age_activity} we describe the derivation of isochronal ages for the entire sample and revisit the age-activity relation. The discussion of our results is presented in Sec. \ref{sec:discussion}. The summary and conclusions are drawn in Sect. \ref{sec:conclusions}.


\section{Data, measurement and calibration} \label{sec:data}
\subsection{Working Sample}

Our sample was selected from the 88 solar twins presented in \citet{ramirez14}.
From this sample, we obtained data for 70 stars with the HARPS instrument \citep{mayor03} at the 3.6 m telescope at the La Silla observatory, to search  for planets around solar twins \citep[program 188.C-0265,][]{bedell15,melendez15,melendez17}. Additional data for 12 stars were found in the ESO archive, as detailed in Table ~\ref{programs}. Thanks to the high quality and cadence time-series observations, combined with the excellent instrumental stability of the HARPS spectrograph, it is possible to explore the limits of chromospheric age-dating using \HK lines.

To measure the solar activity index we used spectra from Ceres, Europa, Vesta, and the Moon (ESO projects 60.A-9036, 60.A-9700, 086.C-0284, 088.C-0323, 092.C-0832, 096.C-0210,289.D-5015,295.C-5031, and 295.C-5035) and correlate with the International Sunspot Number from WDC-SILSO (version 2.0), Royal Observatory of Belgium, Brussels.\footnote{http://www.sidc.be/silso}

\subsection{Contamination of spectroscopic binaries}\label{sec:spec_binary}


We have visual and spectroscopic binaries in our sample, as marked in Table ~\ref{tabelao}. Spectroscopic binaries may have a different evolution from the other stars because, in principle, the interaction with its partner can change the angular momentum and consequently the chromospheric activity, so they were ignored in the age-activity analysis. We cross-matched our sample with the subsample of spectroscopic binaries analysed by \citet{santos17}, \citet{tuccimaia16} and \citet{fuhrmann17}. In addition, we removed the remaining stars with companions within 4'', accourdingly to these studies. In total, 21 stars fell in these selection criteria: HIP6407, HIP14501, HIP18844, HIP19911, HIP30037, HIP54102, HIP54582, HIP62039, HIP64150, HIP64673, HIP65708, HIP67620, HIP72043, HIP73241, HIP79578, HIP81746, HIP83276, HIP87769, HIP103983, HIP109110, and HIP116906. Some of the spectroscopic binaries show enhanced rotation velocities for their ages \citep{santos16,santos17}. Illustrating a few cases, HIP67620 has also been identified as anomalously high in [Y/Mg] \citep{tuccimaia16}, being evidence of mass transfer from a former AGB companion, causing a rejuvenation in stellar activity due to transfer of angular momentum. This star is probably a solar twin blue straggler, like HIP10725 \citep{schirbel15}. The stars HIP19911 also has enhanced Y abundances for its age \citep{tuccimaia16}, suggesting a link to the blue straggler phenomenon.

\subsection{Calibration to the Mount Wilson System}\label{sec:calibmw}

The \HK H and K activity indices were calculated from HARPS spectra following Mount Wilson (MW) prescriptions presented in \citet{wright04}. We compared our $S_{\rm HARPS}$ index for the entire sample of solar twins against their respective \smw\, found in the literature \citep{duncan91, henry96, wright04, melendez09, jenkins11, ramirez14}. In order to provide a more reliable calibration, a subsample of solar twins with the lowest \smw\, uncertainties ($\sigma$ $\leq$ 0.012) were selected, excluding the Sun. From this subsample, our S index measurements were converted into the MW system, resulting in the following transformation equation:
\begin{equation}
\label{eq:pra_s}
\smw=0.9444\,S_{\rm HARPS} + 0.0475,
\end{equation}
where $S_{\rm HARPS}$ is defined as:
\begin{equation}
\label{eq:pra_s2}
S_{\rm HARPS} = 18.349\,\frac{H+K}{R+V}.
\end{equation}
The typical standard deviation for the most inactive stars is 0.004 (\smw $\leq$ 0.190) and 0.014 for the active ones (\smw > 0.190). For each sample star, we provide in Table ~\ref{programs} its \smw\, collected from the literature as well as their respective ESO project identifications. The average values and standard deviation of the S values, already calibrated to the Mount Wilson system using the Eq. ~\ref{eq:pra_s}, are given in Table ~\ref{tabelao}. 

We tested the possibility of \smw\, offsets between the observations performed before and after the HARPS June 2015  upgrade \citep{locurto15}. Considering our \smw\, calibration uncertainties, our results based on 46 stars indicate that both epochs are statistically similar, since the median \smw\, absolute deviation is 0.003, which turns out to be only $\sim$ 1\% of their S values.

\subsection{An improved activity scale for \lrlhk indices}\label{sec:scale_activity}

From the S index we made the conversion to $R_{\rm HK}$, which is the total flux ($F$) in units of erg  cm$^{-2}$s$^{-1}$ at the stellar surface in the H and K lines normalized by the bolometric flux ($F_{\rm HK}/\sigma$T$_{\text{eff}}^{4}$). However R$_{\text{HK}}$ has a strong photospheric contamination (R$_{\text{phot}}$) that needs to be properly corrected in order to pull out the chromospheric signature of the \HK H \& K lines (R$_{\text{HK}}^{'}$). Thus, as a first step, we strictly followed the prescriptions from \citet{wright04} that calibrate the activity measurements as a function of (B-V) color indices and \smw:
\begin{equation}\label{eq:rhk_mw}
R_{\text{HK}} = 1.34 \times 10^{-4}\,C_{\text{cf}}\,\smw,
\end{equation}
where
\begin{equation}\label{eq:ccf_mw}
\log C_{\text{cf}}{\rm (B-V)} = 1.13(B-V)^{3}-3.91(B-V)^{2}+2.84(B-V)-0.47.
\end{equation}
The $C_{\text{cf}}$ term is proportional to the bolometric normalized absolute continuum flux in the R and V Mount Wilson passbands, and the photospheric correction as a function of (B-V) is given by:
\begin{equation}\label{eq:rphot_mw}
\log R_{\text{phot}} {\rm (B-V)}=-4.898+1.918(B-V)^{2}-2.893(B-V)^{3}.
\end{equation}
Finally, we obtained our activity indices $R_{\text{HK}}^{'}$ through Eqs. \ref{eq:rhk_mw}-\ref{eq:ccf_mw}, and then subtracting Eq. \ref{eq:rhk_mw} by Eq. \ref{eq:rphot_mw}:
\begin{equation}
R_{\text{HK}}^{'}=R_{\text{HK}}-R_{\text{phot}}.
\end{equation}

The applicability of these equations are limited to late-F up to early K dwarfs. Recently, \citet{suarez15,suarez16} extended the validity of $C_{\text{cf}}$ and $R_{\text{phot}}$ calibrations towards the M dwarf regime (0.4 $\lesssim$ (B-V) $\lesssim$ 1.9). 
	
In order to test the consistency of our activity measurements, we cross-matched our sample with those of \citet{lovis11} and found 14 solar-twins in common. The mean difference in \lrlhkb\, between both databases (Lovis-Ours) is $\Delta\lrlhkb$  = +0.006 $\pm$ 0.033 dex. The (B-V) colors were taken mostly from the solar twin catalogue of UBV photometry by \citet{ramirez12a} and complemented with other values from the literature, as explained in \citet{ramirez14}. We did not correct for redenning because most of our sample are located within a volume around the centre of a dust-free cavity \citep{lallement14}. The only exception is HIP114615 (d=103$^{+22}_{-15}$ pc) which is at a high galactic latitude (b=-68$^{\circ}$), having thus a negligible extinction, E(B-V)=0.020 according to \citet{schlegel98}, and 0.017 according to the correction by \citet{schlafly11} or 0.008 adopting the correction proposed by \citet{melendez06}.

The equations \ref{eq:ccf_mw} and \ref{eq:rphot_mw} have the disadvantage of using the (B-V) color to calculate the photospheric and chromospheric contributions of \HK H \& K lines, but these are directly related to \Teff\, and \feh\, rather than (B-V) \citep{rochapinto98,lovis11,lorenzo16}. So, in order to minimize these degeneracies, we recalibrated the MW system by replacing the (B-V) for \Teff. To do so, we cross-matched the 72 stars of \citet{noyes84} with those of \citet{ramirez13}. This subsample covers a wide range of \Teff, from 4350 K up to 6500 K. Thus, we investigated a new relation between R$_{\text{phot}}$ \citep[Table 1]{noyes84} and \Teff\, \citep{ramirez13} (plotted in Fig. ~\ref{fig:ccfrphot}):

\begin{figure}[!htbp]
\centering
  \begin{minipage}[t]{0.8 \linewidth}
\centering
    \resizebox{\hsize}{!}{\includegraphics{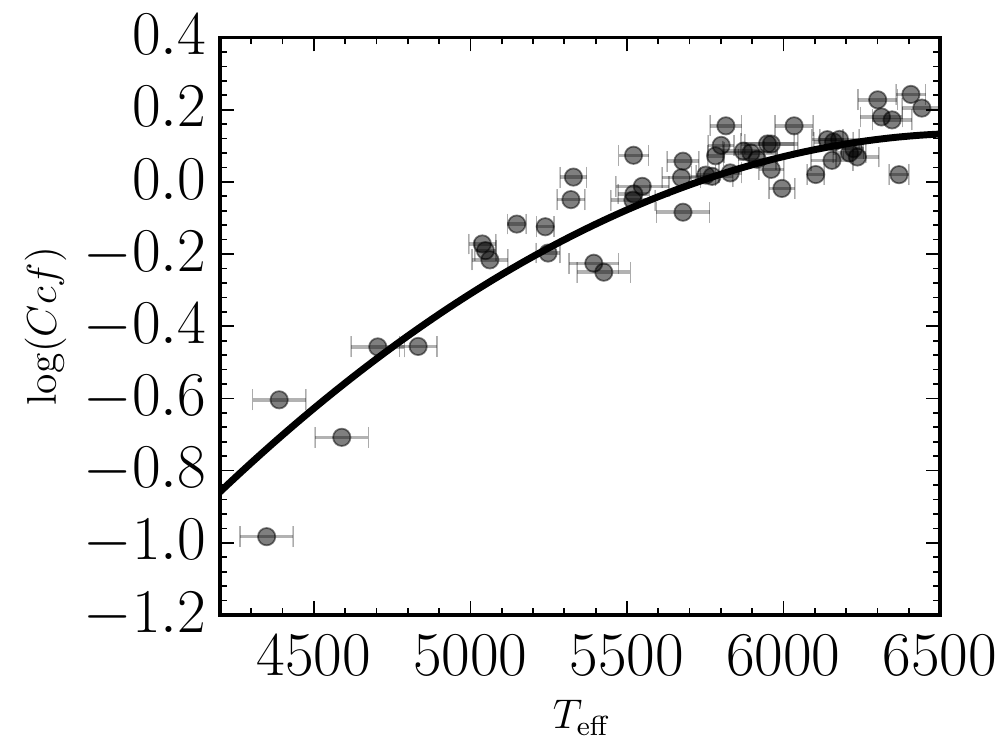}}
  \end{minipage}
  \begin{minipage}[t]{0.8\linewidth}
\centering
    \resizebox{\hsize}{!}{\includegraphics{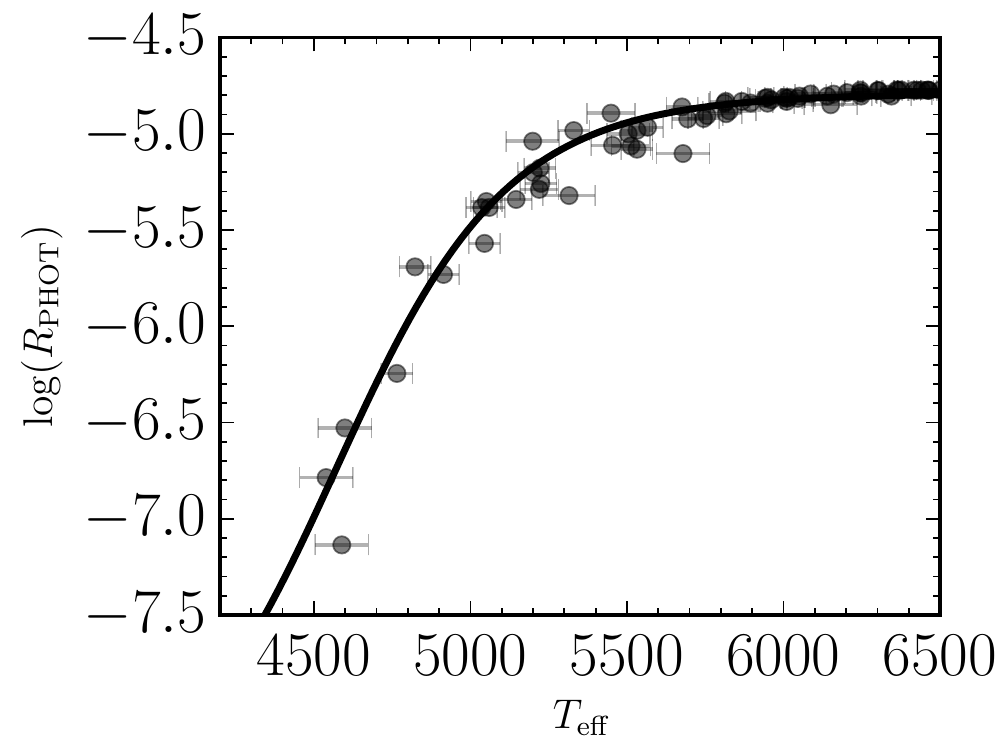}}
  \end{minipage}
  \caption{Upper panel: The C$_{\text{cf}}$ of the stars cited in \citet{rutten84}. The black line represents our fit described in Eq.~\ref{eq:ccf}. Lower panel: The $\log$  $R_{\text{phot}}$ of the stars from \citet{noyes84}. The black line represents our fit described in Eq. ~\ref{eq:rphot}.}
\label{fig:ccfrphot}
\end{figure}

\begin{equation}	
\label{eq:rphot}		
\log R_{\text{phot}} (\Teff)=-4.78845 - \frac{3.70700}{1+(T_{\text{eff}}/4598.92)^{17.5272}}.
\end{equation}

After that, we calibrated the $C_{\text{cf}}$ as a function of \Teff using the values found in \citet{rutten84}, excluding the giant stars. In total, 52 stars in the original \citet{rutten84} sample were cross-matched with those from \citet{ramirez12,ramirez13}. The calibration follows below (plotted in Fig. ~\ref{fig:ccfrphot}):
\begin{equation}
\label{eq:ccf}
\log C_{\text{cf}} (\Teff)=(-1.70\times10^{-7})\,T_{\text{eff}}^2+(2.25\times10^{-3})\,T_{\text{eff}} -7.31.
\end{equation}

\begin{figure}[]
\centering
	\includegraphics[width=1.0\columnwidth]{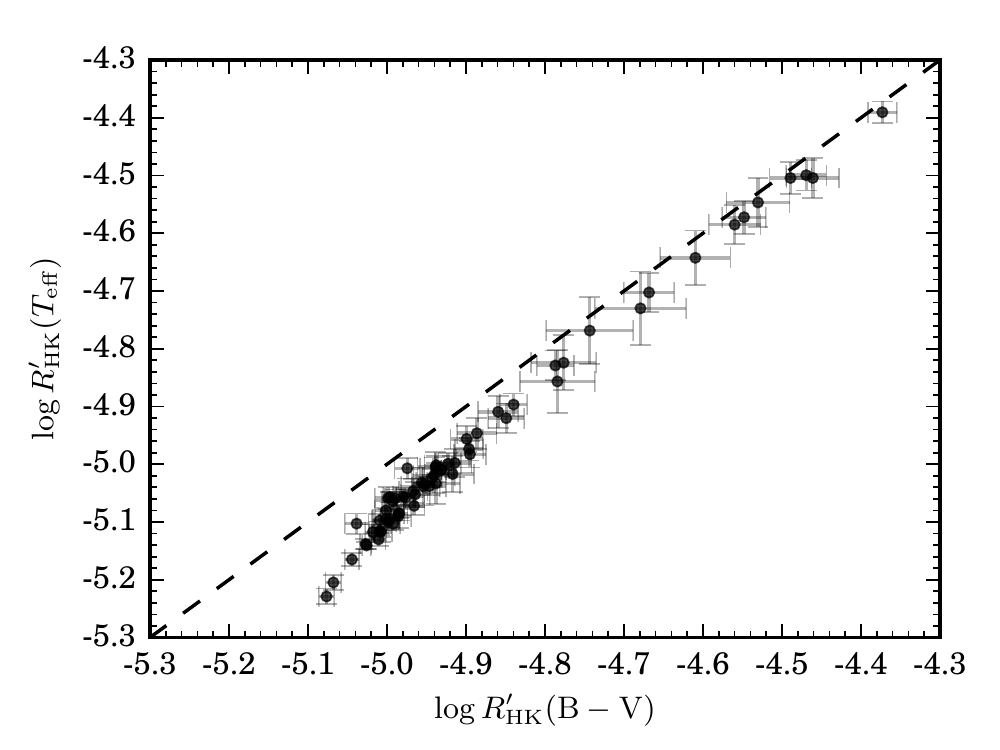}
\caption{The \lrlhk\, using (B-V) vs. the \lrlhk\, using the \Teff. The errorbars represent the intrinsic dispersion of the multiple observations. The black traced line represents the 1:1 relation.}
\label{fig:comp}
\end{figure}

Fig ~\ref{fig:comp} shows that both approaches are strongly correlated to each other. The only significant difference appears as we consider progressively more inactive stars, where the classical MW activity indices seem to decrease their sensitivity to small activity variations, as evidenced by our new approach. Therefore, we expect that the \lrlhkt\, should be a better indicator of activity evolution of most inactive and old stars. Probably, the use of B-V color collapses the effective temperature and metallicity effects which might give more direct information about the absolute continuum flux distribution \citep[see, e.g.,][]{lorenzo16b}. This effect is more evident in inactive stars where the chromospheric/photospheric contrast is weaker.

\subsection{The activity variability of the Sun and solar twins}

\begin{figure}
\centering
	\includegraphics[width=1.0\columnwidth]{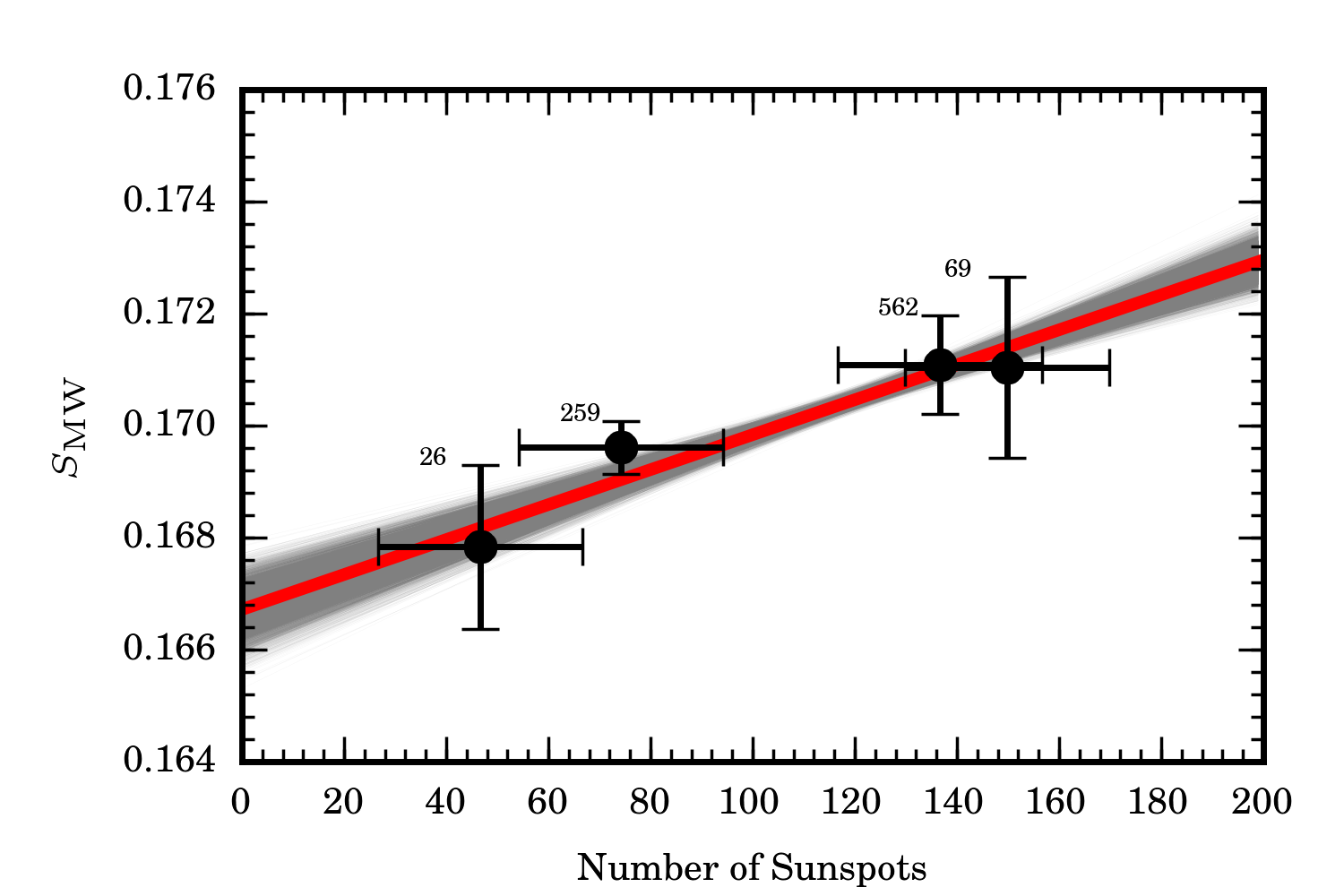}
\caption{The measured S index against the International Sunspot Number (WDC-SILSO) around the same day. The red line represents the best fit between them, as presented in Eq. \ref{eq:spot_s}. The gray lines are the Bisector regression fitting of $10^5$ Monte Carlo simulations based on the activity dispersion in each sunspot number bin. The numbers placed on the top of each error bar represent the number observations that were considered to estimate its mean and dispersion.}
\label{fig:spot}
\end{figure}

\begin{figure}[!htbp]
\centering
  \begin{minipage}[t]{1\linewidth}
\centering
    \resizebox{\hsize}{!}{\includegraphics{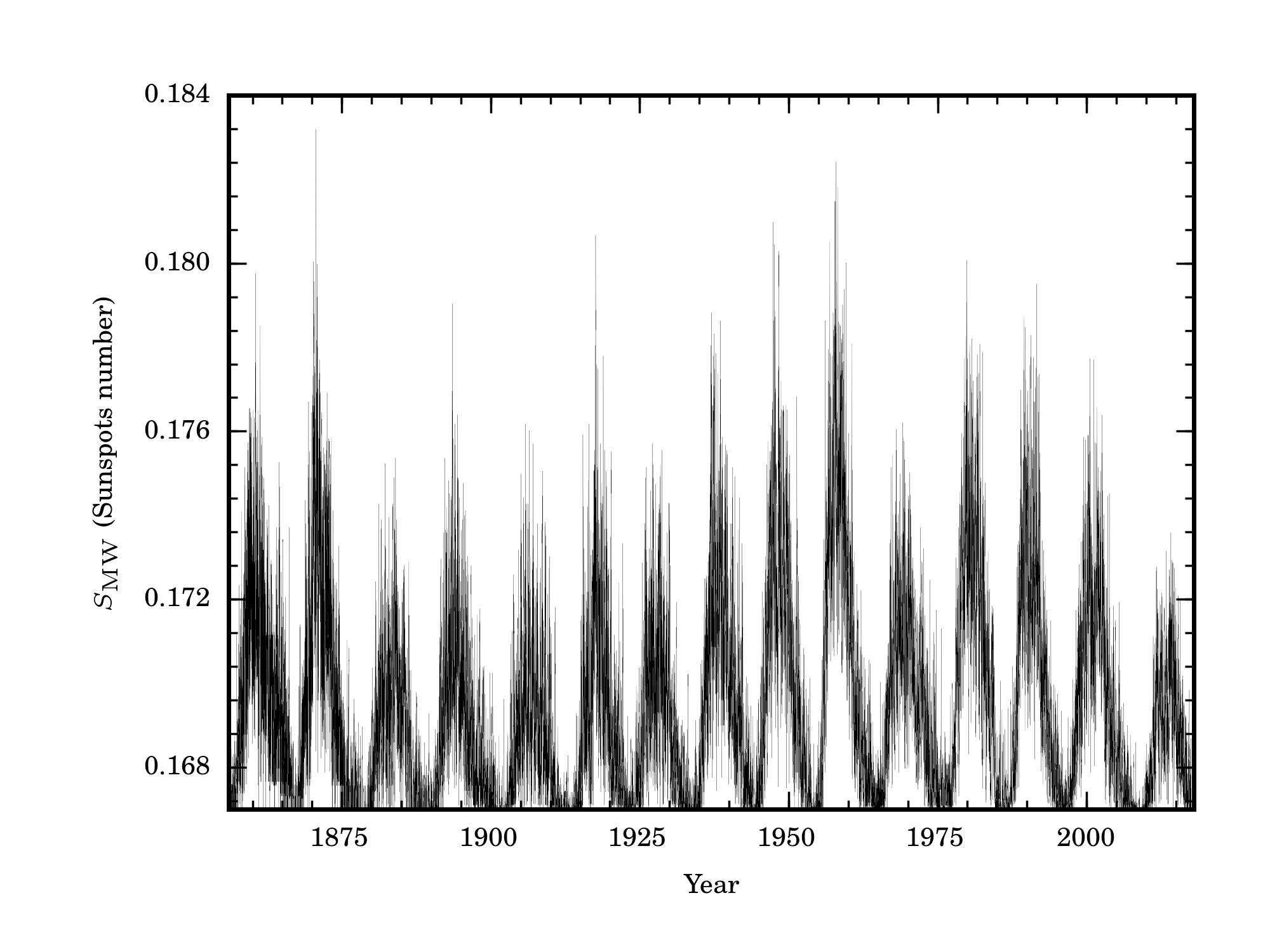}}
  \end{minipage}
    \caption{The daily solar S index (using the eq. ~\ref{eq:spot_s}) since 1850. Solar chromospheric cycles 10$-$24 reconstructed from the relation between \smw vs. Number of sunspots (see Fig. \ref{fig:spot}).
}
\label{fig:cycle1}
\end{figure}

It has been shown by \citet{bertello16} that the disk integrated \HK\,index has a strong linear correlation with sunspot number, although the \HK\, line fluxes are expected to be better correlated to solar plages. In any case, solar plages and the presence of sunspots are different manifestations of the same underlying phenomena (namely magnetic activity) and, therefore, they should be somewhat related to each other. Thus we could perform a calibration between the S-index and sunspot number, allowing us to increase our time baseline, for obtaining a more accurate average S-index. For each day that the S-index was measured we related it with the mean between the number of sunspots one day earlier and one day after the observation, using the WDC-SILSO sunspot numbers. In Fig. \ref{fig:spot}, we show the correlation between our \smw\, and sunspot number. The sunspot number and the activity measurements were binned into 4 intervals of 40 sunspots each with their respective average (in activity and sunspot number) and dispersion represented by the error bars. Through $10^5$ Monte Carlo simulations, assuming gaussian error distribuition, we derived a mean relation between the solar activity and sunspot number, followed by its respective uncertainties:     
\begin{equation}
\label{eq:spot_s}
\smw= (3.12 \pm 0.28) \times 10^{-5}\,N + (0.1667 \pm 0.0003),
\end{equation}
where $N$ is the International Sunspot Number defined by the Royal Observatory of Belgium. The internal error of this approach is $\sigma_{S_{\rm MW}}$ = 0.00038 $\pm$ 0.00009. This relation allowed us to estimate the solar activity level along the cycles 10-24 (1856$-$2017, see Fig. \ref{fig:cycle1}). We found  $<S_{\rm MW}>(10-24)$ = 0.1694 $\pm$ 0.0024  ($\pm$ 0.0004, from Eq. \ref{eq:spot_s}). To check the consistency of our reconstructed solar activity history, we restricted our predictions to cycles 15$-$24, also analysed by \citet{egeland17} who found $<S_{\rm MW}>(15-24)$ = 0.1694 $\pm$ 0.0020. Our result of $<S_{\rm MW}> (15-24)$ = 0.1696 $\pm$ 0.0025 indicates a similar mean activity level and dispersion along these cycles, confirming the overall consistency of our approach. Also, we averaged the cycles 23-24 activity measurements from HARPS observations of the Moon and the other solar-system bodies (Ceres, Vesta, and Europa, hereafter SSB). The differences between them were negligible ($<S_{\rm MW}>^{\rm Moon}$ = 0.1714 $\pm$ 0.0011 and $<S_{\rm MW}>^{\rm SSB}$ = 0.1706 $\pm$ 0.0027) so we combined all available spectra in order to obtain a more consistent measument of the solar activity level ($<S_{\rm MW}>^{\rm SSB + Moon}$ = 0.1712 $\pm$ 0.0017). This result derived from HARPS spectra is also in agreement with our predictions for $<S_{\rm MW}>(10-24)$. For an extensive discussion about \smw\, determinations and calibration issues among different authors and instruments see \citet{egeland17}; in this context our results for the Sun are accurate, as its measurements were made with the same instrumentation as for the stars calibrated into the MW scale. 

\begin{figure}[!htbp]
\centering
   \begin{minipage}[t]{1\linewidth}
\centering
    \resizebox{\hsize}{!}{\includegraphics{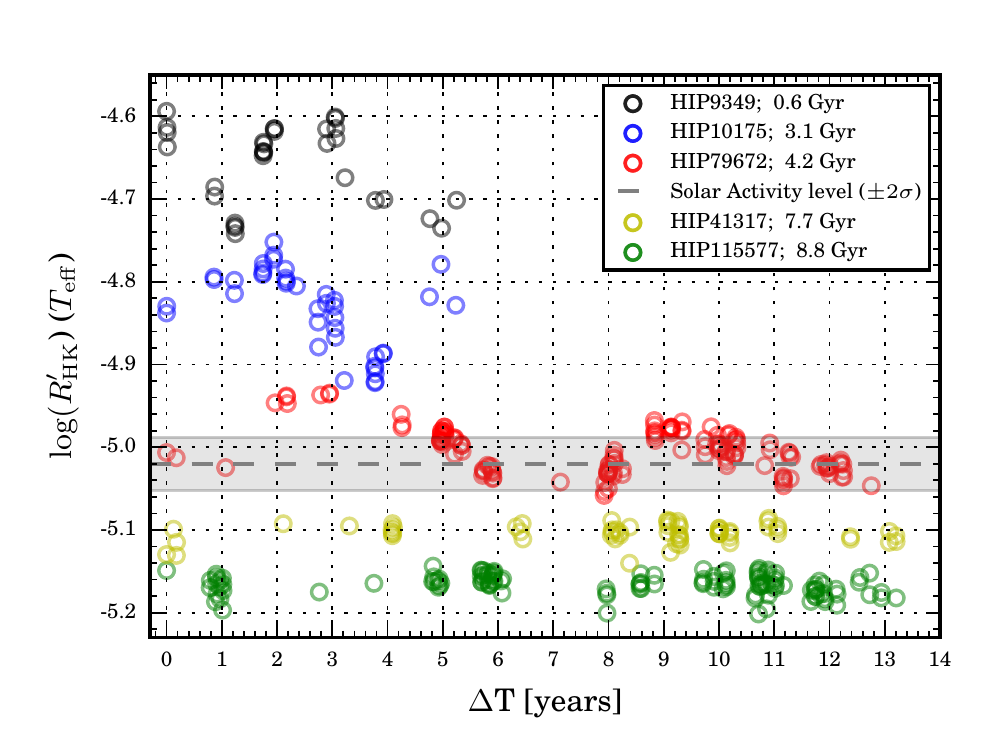}}
  \end{minipage}
  \caption{Solar twins activity modulations over the years of observations. The gray dashed line and filled area correspond to the mean Sun's \lrlhkt\, level and its fluctuations within $\pm$ 2$\sigma$, respectively. Each circle is the nightly averaged level of activity, for multiple observations.
}
\label{fig:cycle2}
\end{figure}

In Fig \ref{fig:cycle2}, a few illustrative cases of stellar chromospheric variability as a function of age are shown. According to our data, the amplitude of activity variations tends to decrease towards older and inactive stars. Interestingly, the well-known solar twin HIP79672 \citep[red circles]{portodemello97,melendez14} shows cycle modulation and amplitude that resembles the Sun (gray shaded region).

\begin{figure}
\centering
	\includegraphics[width=1\columnwidth]{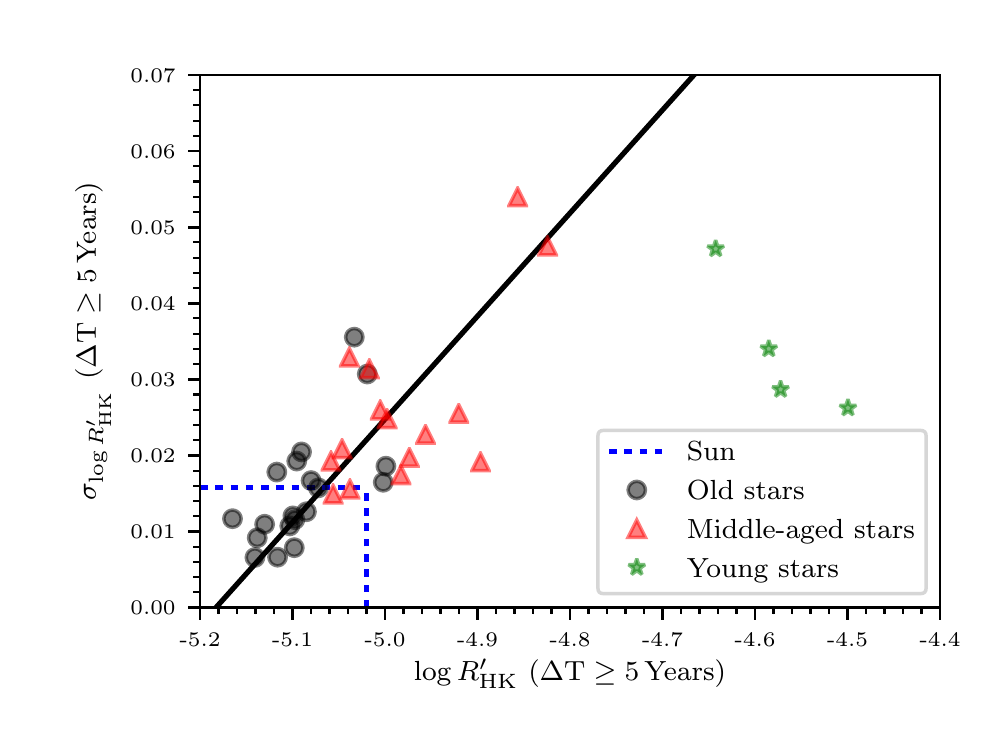}
	\caption{$\sigma_{\log R^\prime_{\rm HK} (T_{\rm eff})}$ vs. \lrlhkt\, relation for stars monitored by more than 5 years. The solid black line is the best fit. Green stars, red triangles, and black circles are solar twins with ages t $\leq$ 2 Gyr, 2 < t $\leq$ 6.5 Gyr, and t $>$ 6.5 Gyr, respectively. The blue dashed line stands for solar activity mean and dispersion.
}
\label{fig:serror}
\end{figure}

It is well known that the amplitude of cycle modulations of cool stars is roughly related to its activity levels \citep{baliunas95,suarez15,suarez16,egeland17} which, for instance, should depend on its evolutionary state \citep{reiners12,schroder13,mittag16}. In this sense, we found in our sample that the standard deviation of \lrlhk\, ($\sigma_{\log(R^\prime_{\rm HK})}$, possibly a proxy of the activity cycle amplitude) due to long-term variations increases towards active stars (Fig. ~\ref{fig:serror}). In order to have a more reliable estimate of $\sigma_{\log(R^\prime_{\rm HK})}$, it is important to monitor the whole activity cycle, however this is likely not the case for most of our sample stars. Thus, we stress that in some cases our derived $\sigma_{\log(R^\prime_{\rm HK})}$ can only represent a lower limit of the realistic $R^\prime_{\rm HK}$ variation during the course of the activity cycles. Even though, we are continously monitoring the cycle modulations of these solar twins over the years and, in the future, we expect to provide a more robust estimate of $\sigma_{\log(R^\prime_{\rm HK})}$ as a function of different activity levels.

In Fig. ~\ref{fig:serror} we separated our stars in three different groups: 1) Young solar twins with ages lower than 2 Gyr were assigned as green stars; 2) Middle-aged solar twins (4.5 $\pm$ 2.0 Gyr, red triangles); 3) Old solar twins with ages greater than 6.5 Gyr (black circles). Stars with time-series observations shorter than 5 years were not considered in order to minimize the effect of short-cycle variations. In the solid black line, we show the linear regression relating mean activity levels \lrlhkt\, and activity dispersion $\sigma_{\lrlhkt}$ fitted to the data. According to our observations, the general trend indicates that the most active stars (\lrlhkt\, $>$ $-$4.7) are in the saturated regime of activity dispersion. Since this region is not well-sampled by our observations, we preferred to rule out these stars from the fit: 
\begin{equation}\label{eq:act_dispersion}
 \sigma_{\lrlhkt} = 0.62 + 0.119 \lrlhkt.
\end{equation}

For instance, considering the measured solar mean activity level of \lrlhkt\, = $-$5.021, the Eq. \ref{eq:act_dispersion} predicts $\sigma_{\log <R^\prime_{\rm HK}>}^{fit} $ = 0.023 which is in agreement with the dispersion measured through cycles 10-24 ($\sigma_{\log <R^\prime_{\rm HK}>}^{\rm cycles\,10-24}$ = 0.016). This result is evidence that the solar variability follows the same trend observed in solar twins. As an example, we applied this equation to the inactive stars shown in Fig.~\ref{fig:cycle2}. Our predictions are in agreement with the observed activity dispersions within 0.005 dex. This relation will be used on Sec. \ref{sec:age_activity} to estimate the lower limits on chromospheric age dating due to cycle variability and also the role of stellar variability on the scatter observed in the age-activity relation.

In Fig. \ref{fig:serror}, it is possible to see that young stars tend to show higher dispersion in their activity measurements while the oldest ones exhibit the lowest activity variations. It is worth noting that the robustness of the activity measurements is a balance between the amount of detectable flux excess (after correting the photospheric signature) and the typical cycle activity modulations. Therefore, younger and older stars tend to show different features in the age-activity diagram. The former shows high levels of activity that can be easily detectable, in contrast with their higher amplitude cycle modulations. In the case of stars with very similar atmospheric parameters, the amplitude of cycle fluctuations can blur minor mass and chemical composition effects on chromospheric indicators. On the other hand, older stars with smaller activity variations and lower flux excess are the most suitable targets to detect these effects on chromospheric indicators. These minor mass and chemical composition effects will be discussed in the next sections.


\section{The activity-age relation} \label{sec:age_activity}
\subsection{Stellar ages}

The isochronal ages of our sample were derived by comparing the observed location of each star in stellar parameter space (\Teff, \logg, \feh, \afe, and V absolute magnitude) with predictions of stellar evolution theory, as computed by the Yonsei-Yale group \citep{yi01,kim02}. Our method is an extension of the procedures adopted in \citet{ramirez14} and \citet{tuccimaia16}, since we included in our analysis other relevant variables that also constrain the morphology of the isochrones such as V magnitude, the trigonometric distance from GAIA DR1 and Hipparcos, and \afe\,information. \footnote{Except for HIP29525 and HIP109110, for which rotational ages were adopted.}. This improved isochronal age-dating approach with additional constraints results in narrower age probability distributions and, consequently, more internally consistent age estimates. Details of this straightforward probabilistic approach are given in \citet{spina17}.

Typically, isochrone ages of main-sequence stars are very uncertain due to poorly-known luminosities, which, in some cases, stems from inaccurate distances/parallaxes, and the fact that stars evolve slowly during that stage. For solar twins, this is not an issue because the stars' precise spectroscopic parameters (\Teff, \feh, \logg, and \afe) are statiscally combined with their luminosities. Indeed, the precision of stellar ages for solar twins is as good as, if not better than those obtained for slightly evolved stars, for which the isochrone method works best. Moreover, because the isochrone sets can be slightly modified to match precisely the solar parameters, ages of solar twins can be made not only very precise, but also reasonably accurate \citep{melendez12,melendez14}. 

In addition, more sophisticated Bayesian approaches to deriving stellar isochronal ages might be necessary to investigate the long-term evolution of heterogeneous populations \citep[see, e.g.,][]{casagrande11}. On the other hand, as we are analysing a sample of stars with very precise atmospheric parameters, the prior distribution becomes approximately constant within the uncertanties of the atmospheric parameters given by the observations \citep{pont04}. In other words, the problem converges to the traditional frequentist chi-squared fit. Moreover, \citet[their Figure 7]{chaname12} have shown that at least one of these approaches which uses Bayesian techniques results in ages which are only slightly offset from those computed using our simpler approach.

Notice that our differential isochrone method gives an age of 4.2$^{+0.3}_{-0.5}$ for 18 Sco, in good agreement with the seismic age of 3.66$^{+0.44}_{-0.50}$ Gyr by \citet{li12}. In addition, for the 16 Cyg pair of solar twins \citep{ramirez11}, our method predicts an age of 6.4$\pm$0.2 Gyr (Tucci-Maia et al. 2018, submitted) which is also close to its seismic age (average of 7.0$\pm$0.1 Gyr) estimated by \citet{vansaders16}. Thus, our method seems valid also for stars around the solar age and somewhat older.

\subsection{Activity-age relation using the updated \lrlhkt}
\label{subsec:teff}

 After averaging all multiple nightly binned activity observations together with its respective standard deviation and estimating the isochronal ages, we analyse now the age-activity diagram of solar twins. The isochronal age-dating method is not optimized for young main-sequence stars. In this region, the isochrones are clumped next to the \textit{Zero Age Main-Sequence}, mapping regimes of very different evolutionary speeds. These differences are translated by a statistical approach into assymetric probability age distributions that are tailed towards older age solutions. This means that, for stars around 1 Gyr, it can be only reasonable to constrain an upper limit for the isochronal ages. Therefore, to overcome this limitation and derive a consistent age-activity relation for younger stars, we chose to simplify our approach assigning a typical age and activity level for this class of stars. Nine stars younger than 1 Gyr (excluding the outlier HIP114615\footnote{This peculiar star was excluded in the analysis because of its very assymetric age errorbar and low activity level.}) were selected in our sample, being classified for the sake of simplicity as a single cluster with mean activity level of \lrlhkt = $-$4.54 $\pm$ 0.09 and a typical age of 0.60$^{+0.19}_{-0.14}$ Gyr which is in good agreement with Hyades' canonical age \cite[0.625 Gyr]{perryman98} and activity level \cite[\lrlhkb = $-$4.50 $\pm$ 0.09]{mamajek08}. It is convenient to stablish $\approx 0.6$ Gyr as our lower limit to young and active stars because, at this age range, according to gyrochronology relations, it is expected the convergence of stellar rotation evolution into well-defined sequence depending only on rotation, age, and mass (or a suitable proxy of it), in a first order approach \citep{barnes07,barnes10,mamajek08}. 

\begin{figure*}[!htbp]
\centering
  \begin{minipage}[t]{0.495 \linewidth}
\centering
    \resizebox{\hsize}{!}{\includegraphics{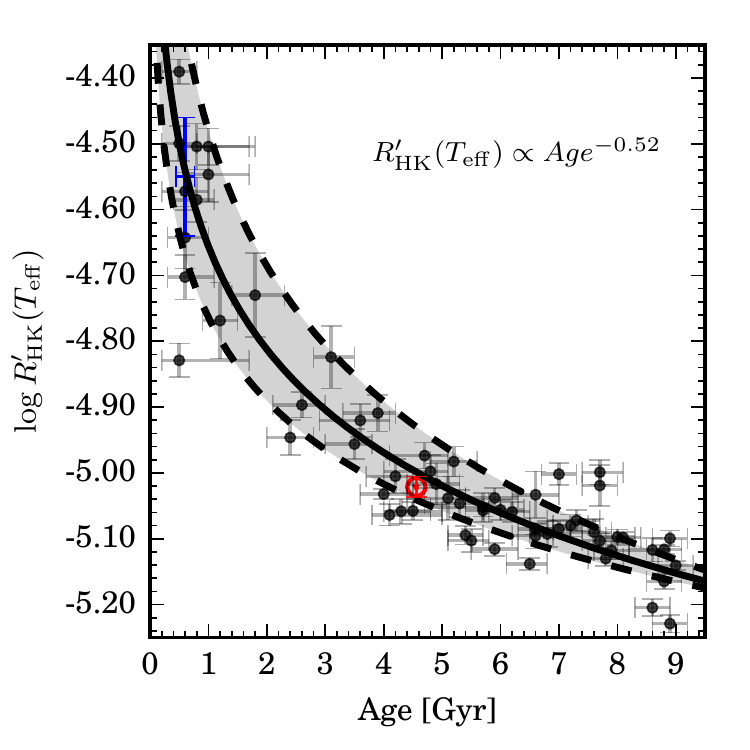}}
  \end{minipage}
  \begin{minipage}[t]{0.495\linewidth}
\centering
    \resizebox{\hsize}{!}{\includegraphics{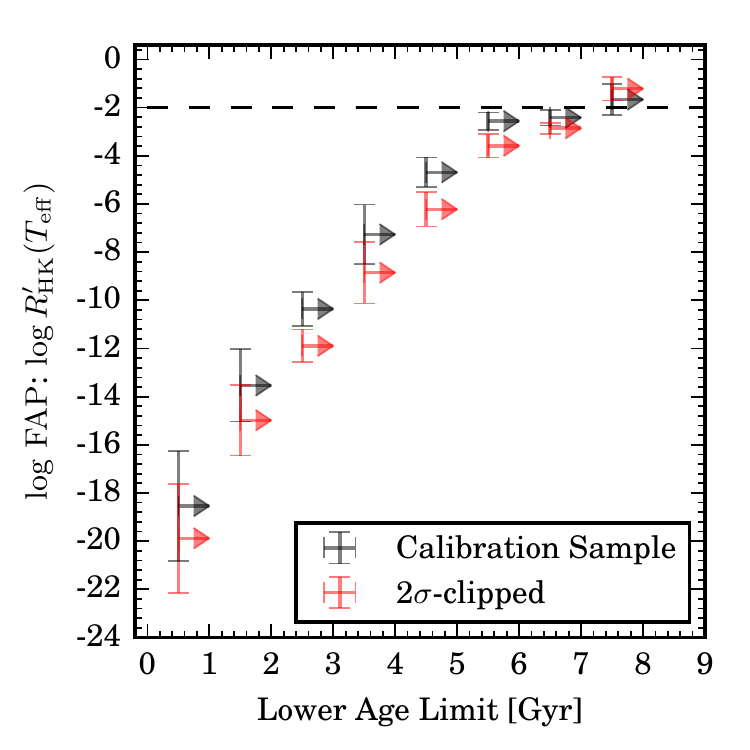}}
  \end{minipage}
  \caption{\textit{Left panel}: Age-Activity relation derived for solar twins. The solid line is the best fit. The Sun is plotted with its usual symbol. Stars younger than 1 Gyr are represented as a single cluster (blue error bar) with mean activity \lrlhkt  = $-$4.54 $\pm$ 0.09 and age = 0.6 $\pm$ 0.2 Gyr. The shaded region is the 2$\sigma$ activity variability prediction band. \textit{Right panel}: Statistical significance of the AC relation as a function of the lower age limit. Black and red symbols are the results for the entire calibration sample and after a single round of 2$\sigma$-clipping removal, respectively. It can be seen that the false-alarm probabilities reach $\approx$ 1\% around 7 Gyr. 
}
\label{fig:ac_fig}
\end{figure*}

In Fig. \ref{fig:ac_fig} (left panel), we show the age-activity relation of solar twins from 0.6 to 9 Gyr. After an extensive radial-velocity monitoring of the whole sample, \citet{santos17} detected a considerable fraction of spectroscopic binaries of 25\% (21 stars, see Sec. \ref{sec:spec_binary}) and an overall multiplicity fraction (taking into account the wide-binary systems) of $\approx$ 42\%. The presence of an unresolved companion in the spectra might bias the determination of atmospheric parameters and, especially, the activity measurements. So, after the RV monitoring, we are confident that our sample of isolated solar twins is suitable for the age-activity (AC) analysis. In the case of wide-binaries, they are visually resolved, showing large orbital separation that prevents the angular momentum transfer between the components, so these targets can be considered as isolated stars. Our age-activity analysis is restricted to a sample of 60 single and wide-binary stars (82 stars $-$ 21 spectroscopic binaries $-$ HIP114615). In addition, the wide-binary star HIP77052 (angular separation of only 4.4'') was also discarded since it shows very assymetric age errorbars, very high level of chromospheric activity for the assigned age, and chemical abundance anomalies reported in \citet{spina17}. As a result, the final sample used to fit the age-activity relation is composed of 59 solar twins spaning the ages from 0.6 to 9 Gyr. In order to posterior check the solar activity behaviour as a function of the other solar twins of same age, we preferred to not include the Sun as an age-activity calibrator. 

Different functional forms were tested to the data and the best solution found was a simple power-law:  
\begin{equation}\label{eq:age_activity}
 \log(Age) = 0.0534 - 1.92\lrlhkt.
 \end{equation}
The error of the slope coefficent is 0.01 and the fractional fitting error found in age is $\approx$ 20\%\footnote{Only two stars in our sample (HIP15527 and HIP44713) show residuals outside the 2$\sigma$ domain predicted by our AC calibration. To evaluate the impact of these stars on AC calibration, we performed a single round of 2$\sigma$-clipping removal and recalibrated the AC relation for the remaining stars. The slope coefficient remained constant (within $\pm$0.01) resulting in identical chomospheric age distributions yielded by both approaches (within $\approx$ 2\%).}. The estimated chromospheric age of the Sun is 4.9 $\pm$ 1.0 Gyr and, for the young cluster of solar twins is 0.63 $\pm$ 0.12 Gyr. It is worth noticing that we are avoiding the young and saturated regime (ages < 0.5 Gyr). So our function is valid for intermediate to old stars (0.6 $\lesssim$ ages $\lesssim$ 9 Gyr) and it can be interpretated as an approximation of a more complex activity evolution that also covers the young and activity saturated regime \citep{mamajek08}. 

It is convenient to estimate the effect of the cycle modulations on the age-activity diagram. So, we propagated the errors of Eq. \ref{eq:age_activity}:
\begin{equation}\label{eq:age_activity2}
 \sigma_{\log(Age)}^{\rm variability} = 1.92\,\sigma_{\log <R^\prime_{\rm HK}>}.
 \end{equation}
 
This equation enable us to estimate the lower limit of chromospheric age error due to stellar cycle variability, assuming that all solar twins follow the age-activity trend shown in Fig. \ref{fig:ac_fig}. The term $\sigma_{\log <R^\prime_{\rm HK}>}$ corresponds to the stellar variability that could be constrained thanks to the multiple observations of our stars (Eq. \ref{eq:act_dispersion}), yielding:
\begin{equation}\label{eq:age_activity3}
 \sigma_{\log(Age)}^{\rm variability} = 1.19 +0.23\lrlhkt.
 \end{equation}

The $\sigma_{\log(Age)}^{\rm variability}$ vs. \lrlhkt\, relation is not well-constrained for stars more active than $\lrlhkt \approx -4.6$.  
In Fig. \ref{fig:ac_fig} (left panel), it is also shown the expected 2$\sigma$ cycle fluctuations following the Eqs. \ref{eq:age_activity} and \ref{eq:age_activity3}. Almost all solar twins are scattered around the overall banana-like trend predicted by our age-activity relation, and the amplitude of the observed scatter is in good agreement with the predicted intrinsic cycle variability, for a given age. 

In order to verify the statistical significance of the age-activity relation, we calculated the Pearson correlation coefficient (R), setting minimum ages starting at 0 and increasing in steps of 0.1 Gyr until 9 Gyr, as shown in Fig.~\ref{fig:ac_fig} (right panel). Then, we binned the age steps in wider intervals of 1 Gyr, estimating the mean false alarm probability and its dispersion within each bin. We can see quantitatively that our data do not follow the age-activity trend found by \citet{pace13}. The false-alarm probability around 2 or 3 Gyr is between $10^{-10}$\% and $10^{-7}$\%, respectively. For stars older than 6-7 Gyr the correlation becomes so low that the probability of a false alarm is greater than 1$\%$. So, in the light of our data, we can confidently say that the age-activity relation evolves until at least 6-7 Gyr. 

On the other hand, no conclusion could be drawn about an \textit{intrinsic lack of activity evolution} after this interval due to poor sampling, age uncertainties and possible influence of other stellar parameters on chromospheric activity levels, for example. 

\begin{figure*}
\centering
  \begin{minipage}[t]{0.495 \linewidth}
\centering
    \resizebox{\hsize}{!}{\includegraphics{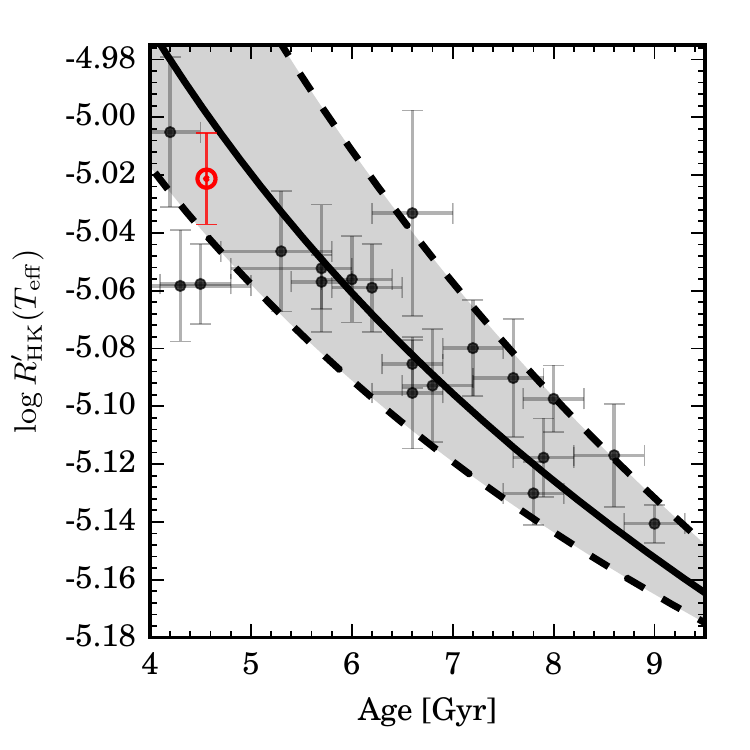}}
  \end{minipage}
  \begin{minipage}[t]{0.495\linewidth}
\centering
    \resizebox{\hsize}{!}{\includegraphics{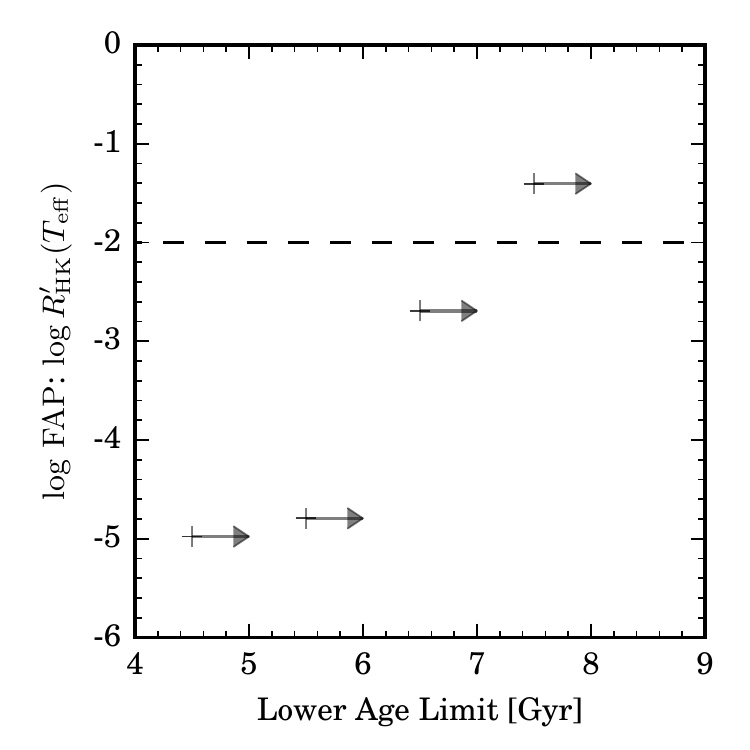}}
  \end{minipage}
  \begin{minipage}[t]{0.495 \linewidth}
  	\centering
  	\resizebox{\hsize}{!}{\includegraphics{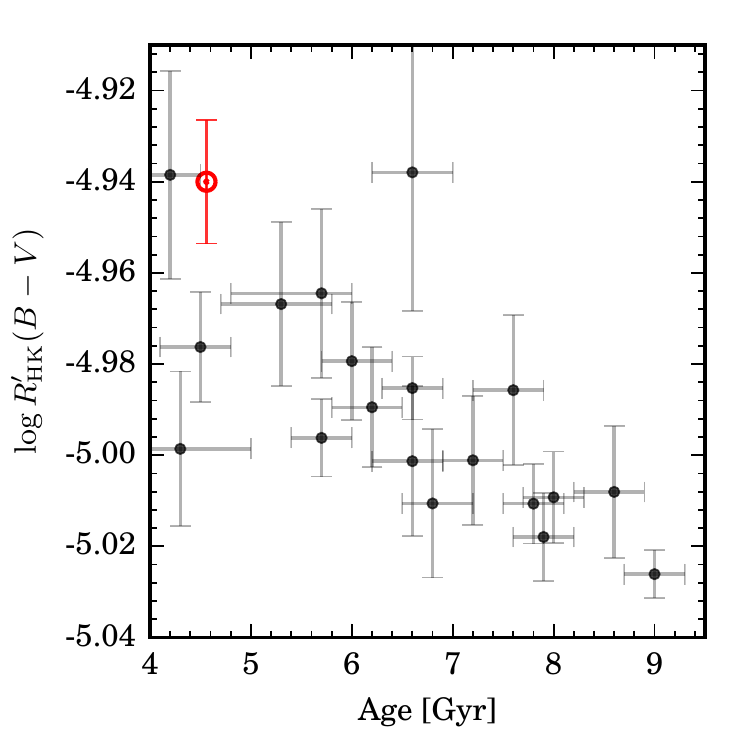}}
  \end{minipage}
  \begin{minipage}[t]{0.495\linewidth}
  	\centering
  	\resizebox{\hsize}{!}{\includegraphics{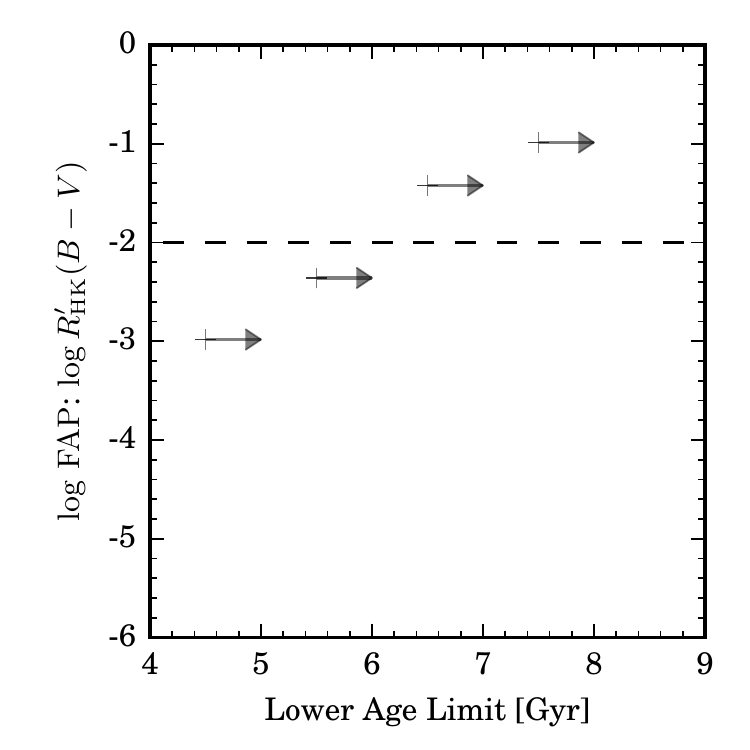}}
  \end{minipage}

	\caption{\textit{Left panels}: End of the age-activity relation of solar twing for \lrlhkt\,(upper panel) and \lrlhkb\,(lower panel). The solid black line is the AC calibration from Eq. \ref{eq:age_activity} and the shaded area represents the 2$\sigma$ variability prediction band for \lrlhkt. The Sun is plotted in red as its usual symbol. \textit{Right panels}: Same statistical analysis of Fig. ~\ref{fig:ac_fig} applied to the best old solar twins available in our sample. 
}
\label{fig:ac_old_suns}
\end{figure*}

Still, we could go one step further and visually inspect in detail the end of the age-activity diagram isolating the variables that are known to affect the activity levels of the most inactive stars such as mass, metallicity and \ion{Ca}{} abundances. Thus, in order to visualize better the pure effect of the age-activity correlation, we restricted our sample to the best old solar twins (age > 4 Gyr) available in our sample within $\pm$ 0.05 of the solar values in $M/M_{\odot}$, \feh, and \textrm{[Ca/H]} \citep{spina17}. In Fig. \ref{fig:ac_old_suns} (upper left panel), the end of the AC diagram is shown, followed by the predictions of Eq. \ref{eq:age_activity} and \ref{eq:age_activity3} for \lrlhkt\, activity index. The upper right panel of Fig. \ref{fig:ac_old_suns} is the same statistical analysis of Fig. ~\ref{fig:ac_fig} (right panel) applied only to the best old solar twins. The same statistical analysis was also repeated for \lrlhkb\, activity index (lower panels) and, albeit with slightly lower statistical significance in comparison to the \lrlhkt\, vs. age analysis, it is still possible to detect the activity evolution until $\approx$ 6 Gyr. 

We confirmed that the AC relation remains statistically relevant after the solar age also for the most homogeneous group of stars. The typical chromospheric age error derived for these stars is $\approx$ 13\% or about 1 Gyr for a typical 7 Gyr old solar twin.  Possibly, the poor sampling after $\approx$ 7 Gyr together with the increasing ratio between isochronal age errors and the dynamical age range (from 7 to 9 Gyr) are responsable for the lack of statistical significance observed after this domain. In Table 3, we show the performance of our age-activity calibration for old stars (age > 1 Gyr) with progressively longer time-span coverage  of Ca II observations (from 5 to $\sim$13 years). Typically, our calibration yields a chromospheric age error of about 15\%, independent on time-span restrictions.

The Sun is a key target to constrain the age-activity relations, therefore it is important to verify whether it has a typical level and dispersion of chromospheric activity in comparison to other stars with similar parameters. It can be seen in Fig. ~\ref{fig:ac_fig} (left panel) and Fig. ~\ref{fig:ac_old_suns} (left panel) that the Sun is a normal star in comparison to other solar twins, following the overall age-activity trend and also a compatible activity dispersion expected for a typical 4 Gyr old star.

\section{Discussion}\label{sec:discussion}

According to our results, the chromospheric activity index \lrlhkt\, is an interesting clock up to $\approx$ 7 Gyr. Besides the age evolution, the chromospheric activity is well-known to be also correlated with stellar mass and other atmospheric parameters such as metallicity. In this regard, our analysis using a group of stars with precise and very similar atmospheric parameters enabled us to mitigate these effects, testing the limits of the age-activity relation. Our analysis does not indicate that the age-activity relation flattens out for stars older than 1 to 3 Gyr as, described by \citet{pace04} and \citet{pace13}. Probably, the simplest explanation for the lack of chromospheric evolution in old field stars is the combination of mass/metallicity dependencies that arises from selection effects \citep{lorenzo16}. In addition to mass effects, distant open clusters members also suffer from interstellar contamination of \HK lines, biasing the activity measurements towards inactive levels \citep{curtis17}. 

If there is a sudden decrease of the activity level followed by a relatively constant and inactive phase, the Rossby number vs. activity diagram should reveal this behaviour. In this sense, \citet{mamajek08} did not find any discontinuity in the age-rotation-activity relation for stars with ages $\gtrsim$ 1 Gyr. Also, in the light of new open cluster data from the Kepler mission, the gyrochronology relations show very consistent results for solar-type stars from 1 to 4 Gyr old \citep{meibom11,meibom15,barnes16}. After $\approx$ 4 Gyr, the scenario of smooth rotational evolution predicted by previous gyrochronology relations is still under debate. \citet{vansaders16} combined the rotational periods of intermediate age open clusters members and old field solar-type stars with measured rotational periods (photometric or asteroseismic rotational periods) to point out that, at some threshold Rossby number, a rapid change in the topology of the magnetic fields happens inside the star and this effect is translated into an inefficient magnetic braking for relatively old stars. In brief, the observational effect of this change is the overabundance of unexpected old rapid rotators. Therefore, according to their analysis, in the case of solar mass stars, for example, the usual gyrochronology relations become ineffective for ages older than $\approx$ 4 Gyr. 

In contrast, our age-activity diagram does not indicate any sign of discontinuity or overpresence of old and relatively rapid rotators due to an inefficient magnetic braking, assuming that activity and rotational evolutions are coupled. \citet{barnes16b} analysed the Kepler sample stars with measured rotational periods and asteroseismic ages and, after removing the metal-poor and post main-sequence stars (\logg $<$ 4.2), they found a good agreement between seismic and rotational ages up to 8-9 Gyr. A similar result was previously found by \citet{donascimento14} analysing solar analogs and candidates of solar twins from the Kepler mission. Both results are consistent with our findings in this work.

\section{Summary and Conclusions}\label{sec:conclusions}

The main goal of this paper is to revisit the activity-age relation using HARPS high-resolution time-series observations of 82 solar twins whose precise isochronal ages and other important physical parameters (such as \Teff, \feh, \logg\, and [Ca/H] abundances) have been obtained \citep{spina17,bedell18}. To do so, the \HK H \& K $S$ indices were calculated following Mount Wilson prescriptions presented in \citet{wright04} and then, we revisited the MW calibration equations to build a new activity index \lrlhkt, replacing the color index dependency by \Teff\, (Eqs. ~\ref{eq:ccf} and ~\ref{eq:rphot}). This modification mitigates the metallicity degeneracy present in (B-V) color indices. 

The solar \smw\, were calculated also from HARPS observations and related to sunspot number (Eq. ~\ref{eq:spot_s}). Thus, anchored on the sunspot number time-series from the Royal Observatory of Belgium, we could reconstruct the solar activity level along the cycles 10-24 (1856-2017, $<S_{\rm MW}>(10-24)$ = 0.1694 $\pm$ 0.0024, which is in excellent agreement with the \citet{egeland17} analysis recalibrating the Sun's activity level. Through multiple observations of solar twins, we detected that younger stars tend to show higher activity dispersion in comparison to older counterparts. Therefore, a simple relation between mean activity level ($<R^\prime_{\rm HK}>$) and long-term activity variation ($\sigma_{<R^\prime_{\rm HK}>}$) could be derived (Eq. ~\ref{eq:act_dispersion}). The solar long-term activity variation follows the same trend observed for solar twins with similar age and mean activity levels, thus we conclude that the Sun has a mean activity level typical for its age. This relation helped us to predict the scatter due to stellar variability on the age-activity evolution of solar twins. 

Interestingly, the age-activity relation found for solar twins follows the Skumanich-like function $<R^\prime_{\rm HK}> \propto Age^{-0.52}$ (Eq. ~\ref{eq:age_activity}) similar to the power-law derived by \citet{soderblom91}. The fractional age uncertainty is around 20\% and the AC relation is valid only for solar-mass solar-metallicity stars with ages between 0.6 to 9 Gyr. Almost all stars in our sample are placed within the predicted AC variability band for a given age, indicating that, in principle, a significant part of the observed scatter could be explained by long-term cycle modulations. For our sample, tests of statistical significance of the age-activity relation rule out the lack of evolution scenario after $\approx$ 2 Gyr proposed by \citet{pace04} and \citet{pace13}. So, our approach can be applied to age-date solar twins to at least 6-7 Gyr, where the false-alarm probability reaches $\approx$ 1\%. Alternatively, as we consider only the best solar twins available in our sample (solar within $\pm$ 0.05 in $M/M_{\odot}$, \feh, and \textrm{[Ca/H]}), the chromospheric activity seems to evolve monotonically towards the end of the main-sequence ($\approx$ 9 Gyr). This result is in line with previous works using open clusters and field stars with precise ages \citep{mamajek08,donascimento14,barnes16b,lorenzo16b}, reinforcing the use of chromospheric activity as an age diagnostic over a wide range of ages. 

\begin{acknowledgements}

We would like to acknowledge the anonymous referee, whose comments have unquestionably led to an improved paper. DLO acknowledges the support from FAPESP (2016/20667-8). JM thanks support from FAPESP (2012/24392-2) and CNPq (Productivity Fellowship).

\end{acknowledgements}

\bibliographystyle{aa} 
\bibliography{bibli2} 

\newpage

%

\newgeometry{left=2.5cm,bottom=2cm}
\small
\onecolumn
\newpage

\Online

\setcounter{table}{0}

\begin{longtab}
{\scriptsize
\setlength\LTleft{-20pt}
\setlength\LTright{0pt}
\begin{longtable}{@{\extracolsep{\fill}}clccl@{}}
\caption{\label{programs} All the solar twins observed by HARPS, their respective programs and $S$ values compiled from the literature.}\\
\hline \hline
HIP                          & \multicolumn{1}{c}{ESO programs}                                                            & S      & $\sigma_{S}$  & \multicolumn{1}{c}{reference} \\
\hline
\endfirsthead
\caption{continued.}\\
\hline \hline
HIP                          & \multicolumn{1}{c}{ESO programs}                                                            & S      & $\sigma_{S}$  & \multicolumn{1}{c}{reference} \\
\hline
\endhead
\hline
\endfoot
1954 & 188.C-0265; 072.C-0488; 096.C-0499; 192.C-0852; 091.C-0936; 183.C-0972; 0100.D-0444  & 0.179 & 0.007 &  1; 3  \\ 
3203 &  188.C-0265; 0100.D-0444  & 0.299 & 0.006 &  1; 4  \\ 
4909 &  188.C-0265; 0100.D-0444  & 0.297 & 0.012 & 1  \\ 
5301 &  092.C-0721; 183.C-0972; 090.C-0421; 072.C-0488; 091.C-0034; 0100.D-0444  & 0.167 & 0.012 & 1  \\ 
6407* &  188.C-0265  & 0.214 & 0.012 & 1  \\ 
7585 &  188.C-0265; 0100.D-0444  & 0.176 & 0.006 &  1; 4  \\ 
8507 &  188.C-0265; 0100.D-0444  & 0.171 & 0.008 &  1; 2  \\ 
9349 &  188.C-0265; 0100.D-0444  & 0.283 & 0.012 & 1  \\ 
10175 &  188.C-0265; 0100.D-0444  & 0.209 & 0.012 & 1  \\ 
10303 &  188.C-0265; 0100.D-0444  & 0.167 & 0.006 &  1; 4  \\ 
11915 &  188.C-0265; 092.C-0721; 093.C-0409; 0100.D-0444  & 0.181 & 0.012 & 1  \\ 
14501* &  188.C-0265; 183.C-0972; 072.C-0488; 192.C-0852; 0100.D-0444  & 0.157 & 0.006 &  1; 4  \\ 
14614 &  188.C-0265; 076.C-0155; 0100.D-0444  & 0.168 & 0.012 & 1  \\ 
15527 &  183.C-0972; 072.C-0488; 192.C-0852; 0100.D-0444  & 0.177 & 0.013 &  1; 3  \\ 
18844* &  188.C-0265  & 0.167 & 0.007 &  1; 3; 5  \\ 
19911* &  188.C-0265  & 0.253 & 0.008 &  1; 4  \\ 
22263 &  188.C-0265; 0100.D-0444  & 0.279 & 0.014 &  1; 2; 4; 5; 6  \\ 
25670 &  188.C-0265; 097.C-0571; 0100.D-0444  & 0.168 & 0.015 &  1; 2  \\ 
28066 &  188.C-0265; 0100.D-0444  & 0.155 & 0.006 &  1; 4; 6  \\ 
29432 &  188.C-0265; 185.D-0056; 0100.D-0444  & 0.168 & 0.006 &  1; 4  \\ 
29525 &  072.C-0488; 0100.D-0444  & 0.343 & 0.012 & 1  \\ 
30037* &  188.C-0265  & 0.166 & 0.008 &  1; 2; 7  \\ 
30158 &  188.C-0265  & 0.169 & 0.011 &  1; 5  \\ 
30476 &  183.C-0972; 188.C-0265; 072.C-0488  & 0.159 & 0.006 &  1; 5  \\ 
30502 &  188.C-0265; 183.D-0729  & 0.168 & 0.006 &  1; 7  \\ 
33094 &  072.C-0488  & 0.153 & 0.012 & 1  \\ 
34511 &  188.C-0265  & 0.166 & 0.012 & 1  \\ 
36512 &  183.C-0972; 188.C-0265; 072.C-0488; 091.C-0936  & 0.170 & 0.006 &  1; 7  \\ 
36515 &  192.C-0224; 0100.D-0444  & 0.363 & 0.012 & 1  \\ 
38072 &  188.C-0265  & 0.305 & 0.014 &  1; 2  \\ 
40133 &  188.C-0265  & 0.160 & 0.006 &  1; 4  \\ 
41317 &  188.C-0265; 183.C-0972; 072.C-0488  & 0.164 & 0.006 &  1; 4; 5; 7   \\ 
42333 &  188.C-0265  & 0.304 & 0.015 &  1; 4  \\ 
43297 &  188.C-0265  & 0.256 & 0.007 &  1; 4  \\ 
44713 &  072.C-0488; 192.C-0852; 196.C-1006  & 0.185 & 0.023 &  1; 3  \\ 
44935 &  188.C-0265; 183.D-0729  & 0.165 & 0.006 &  1; 7  \\ 
44997 &  183.D-0729; 188.C-0265; 075.C-0332  & 0.174 & 0.012 &  1; 7  \\ 
49756 &  183.D-0729; 188.C-0265  & 0.164 & 0.006 &  1; 4  \\ 
54102* &  188.C-0265; 072.C-0488  & 0.218 & 0.008 &  1; 7  \\ 
54287 &  183.D-0729; 188.C-0265; 183.C-0972; 072.C-0488  & 0.192 & 0.046 &  1; 5  \\ 
54582* &  183.D-0729; 072.C-0488; 183.C-0972; 188.C-0265  & 0.155 & 0.006 &  1; 4  \\ 
62039* &  188.C-0265; 183.D-0729  & 0.155 & 0.006 &  1; 4  \\ 
64150* &  188.C-0265  & 0.159 & 0.006 &  1; 4; 6  \\ 
64673 &  188.C-0265  & 0.163 & 0.012 & 1  \\ 
64713 &  188.C-0265; 183.D-0729  & 0.167 & 0.006 &  1; 7  \\ 
65708* &  188.C-0265  & 0.155 & 0.006 &  1; 4  \\ 
67620* &  188.C-0265  & 0.215 & 0.008 &  1; 4; 5  \\ 
68468 &  188.C-0265; 097.C-0571  & 0.156 & 0.012 & 1  \\ 
69645 &  188.C-0265  & 0.164 & 0.012 & 1  \\ 
72043* &  188.C-0265  & 0.168 & 0.006 &  1; 4  \\ 
73241* &  188.C-0265  & 0.172 & 0.014 &  1; 5  \\ 
73815 &  188.C-0265; 075.C-0332  & 0.161 & 0.012 & 1  \\ 
74389 &  072.C-0488  & 0.171 & 0.012 & 1  \\ 
74432 &  188.C-0265  & 0.149 & 0.006 &  1; 4  \\ 
76114 &  188.C-0265  & 0.161 & 0.006 &  1; 4  \\ 
77052 &  075.C-0332; 188.C-0265  & 0.196 & 0.006 &  1; 4; 5  \\ 
77883 &  188.C-0265; 183.D-0729  & 0.166 & 0.006 &  1; 7  \\ 
79578* &  188.C-0265  & 0.217 & 0.006 &  1; 5  \\ 
79672 &  183.D-0729; 185.D-0056; 188.C-0265; 192.C-0852; 183.C-0972; 072.C-0488; 196.C-1006; 077.C-0364; 099.C-0491  & 0.170 & 0.006 &  1; 2; 4; 6; 7  \\ 
79715 &  188.C-0265  & 0.172 & 0.013 &  1; 5  \\ 
81746* &  188.C-0265  & 0.173 & 0.011 &  1; 5  \\ 
83276* &  188.C-0265  & -- & -- & --  \\ 
85042 &  188.C-0265; 183.C-0972; 192.C-0852; 089.C-0415; 072.C-0488  & 0.157 & 0.006 &  1; 4  \\ 
87769* &  188.C-0265  & 0.170 & 0.012 & 1  \\ 
89650 &  188.C-0265; 183.D-0729; 0100.D-0444  & 0.165 & 0.006 &  1; 7  \\ 
95962 &  188.C-0265; 183.C-0972; 072.C-0488; 60.A-9036; 077.C-0364; 0100.D-0444  & 0.161 & 0.006 &  1; 4  \\ 
96160 &  188.C-0265; 0100.D-0444  & 0.187 & 0.012 & 1  \\ 
101905 &  188.C-0265; 183.D-0729; 0100.D-0444; 099.C-0491  & 0.216 & 0.019 &  1; 3; 3; 5   \\ 
102040 &  188.C-0265; 183.D-0729; 0100.D-0444  & 0.170 & 0.006 &  1; 4; 6  \\ 
102152 &  292.C-5004; 188.C-0265; 183.D-0729; 0100.D-0444  & 0.161 & 0.012 & 1  \\ 
103983* &  188.C-0265  & -- & -- & --  \\ 
104045 &  188.C-0265; 097.C-0571; 092.C-0721; 093.C-0409; 0100.D-0444  & 0.164 & 0.012 & 1  \\ 
105184 &  188.C-0265; 183.D-0729; 0100.D-0444  & 0.231 & 0.022 &  1; 3; 5  \\ 
108158 &  072.C-0488; 183.C-0972; 0100.D-0444  & -- & -- & --  \\ 
108468 &  188.C-0265; 072.C-0488; 183.D-0729; 183.C-0972; 091.C-0936; 192.C-0852; 0100.D-0444  & 0.163 & 0.012 & 1  \\ 
109821 &  192.C-0852; 072.C-0488; 196.C-1006; 0100.D-0444  & 0.159 & 0.012 & 1  \\ 
114328 &  188.C-0265; 0100.D-0444  & -- & -- & --  \\ 
114615 &  188.C-0265  & 0.191 & 0.012 & 1  \\ 
115577 &  072.C-0488; 188.C-0265; 192.C-0852; 183.C-0972; 183.D-0729; 196.C-1006; 0100.D-0444  & 0.160 & 0.006 & 1; 5  \\ 
116906* &  072.C-0488; 183.C-0972; 192.C-0852  & 0.163 & 0.012 & 1  \\ 
117367 &  188.C-0265; 0100.D-0444  & 0.156 & 0.012 & 1  \\ 
118115 &  188.C-0265; 0100.D-0444  & 0.16 & 0.012 & 1  \\ 
\hline                                                                    
\end{longtable}
\begin{longtable}{l}
1 \citet{ramirez14} \\
2 \citet{jenkins11} \\
3 \citet{jenkins06} \\
4 \citet{wright04} \\
5 \citet{henry96} \\
6 \citet{duncan91} \\
7 \citet{melendez09}
\end{longtable}}
\end{longtab}
\newpage
\setcounter{table}{1}
\begin{longtab}
{\scriptsize
\setlength\LTleft{-20pt}
\setlength\LTright{0pt}
\begin{longtable}{@{\extracolsep{\fill}}lcccccccccc@{}}
\caption{\label{tabelao} Parameters of all sample solar twins.}\\
\hline \hline
HIP & V & (B-V) & \Teff & \logg & \feh & $S_{\rm MW}$ & \lrlhkt  & age (Gyr)  & N$^{\rm o}$ Obs & time-span (years) \\
\hline
\endfirsthead
\caption{continued.}\\
\hline \hline
HIP & V & (B-V) & \Teff & \logg & \feh & $S_{\rm MW}$ & \lrlhkt  & age (Gyr)  & N$^{\rm o}$ Obs & time-span (years) \\
\hline
\endhead
\hline
\endfoot

1954 & 7.275 $\pm$ 0.004 & 0.681 $\pm$ 0.007 & 5720 $\pm$ 2 & 4.460 $\pm$ 0.008 & -0.090 $\pm$ 0.003 & 0.177 $\pm$ 0.004 & -5.000 $\pm$ 0.025 &    4.8$^{+0.3}_{-0.8}$    & 87 &    14.0  \\ 
3203 & 7.030 $\pm$ 0.008 & 0.620 $\pm$ 0.015 & 5868 $\pm$ 9 & 4.540 $\pm$ 0.016 & -0.050 $\pm$ 0.007 & 0.314 $\pm$ 0.013 & -4.500 $\pm$ 0.027 &    0.5$^{+0.3}_{-0.3}$    & 35 &    6.0  \\ 
4909 & 8.512 $\pm$ 0.006 & 0.637 $\pm$ 0.024 & 5861 $\pm$ 7 & 4.50 $\pm$ 0.016 & 0.048 $\pm$ 0.006 & 0.282 $\pm$ 0.012 & -4.572 $\pm$ 0.028 &    0.6$^{+0.4}_{-0.4}$    & 34 &    6.0  \\ 
5301 & 8.449 $\pm$ 0.012 & 0.650 $\pm$ 0.009 & 5723 $\pm$ 3 & 4.395 $\pm$ 0.011 & -0.074 $\pm$ 0.003 & 0.165 $\pm$ 0.002 & -5.074 $\pm$ 0.016 &    7.3$^{+0.4}_{-0.5}$    & 10 &    11.2  \\ 
6407*   & 8.624 $\pm$ 0.002 & 0.652 $\pm$ 0.015 & 5775 $\pm$ 7 & 4.505 $\pm$ 0.013 & -0.058 $\pm$ 0.006 & 0.223 $\pm$ 0.004 & -4.772 $\pm$ 0.015 &    1.9$^{+0.7}_{-0.7}$    & 7 &    1.2  \\ 
7585 & 6.764 $\pm$ 0.007 & 0.648 $\pm$ 0.008 & 5822 $\pm$ 3 & 4.445 $\pm$ 0.008 & 0.083 $\pm$ 0.003 & 0.178 $\pm$ 0.004 & -4.955 $\pm$ 0.023 &    3.5$^{+0.3}_{-0.5}$    & 81 &    6.0  \\ 
8507 & 8.899 $\pm$ 0.004 & 0.651 $\pm$ 0.018 & 5717 $\pm$ 3 & 4.460 $\pm$ 0.011 & -0.099 $\pm$ 0.003 & 0.174 $\pm$ 0.005 & -5.016 $\pm$ 0.030 &    4.9$^{+0.4}_{-0.5}$    & 44 &    6.0  \\ 
9349 & 7.992 $\pm$ 0.017 & 0.650 $\pm$ 0.009 & 5818 $\pm$ 6 & 4.515 $\pm$ 0.011 & -0.006 $\pm$ 0.005 & 0.260 $\pm$ 0.017 & -4.643 $\pm$ 0.048 &    0.6$^{+0.4}_{-0.3}$    & 33 &    6.0  \\ 
10175 & 8.18 $\pm$ 0.016 & 0.704 $\pm$ 0.018 & 5719 $\pm$ 3 & 4.485 $\pm$ 0.010 & -0.028 $\pm$ 0.002 & 0.217 $\pm$ 0.014 & -4.815 $\pm$ 0.053 &    3.1$^{+0.4}_{-0.3}$    & 52 &    6.0  \\ 
10303 & 7.629 $\pm$ 0.013 & 0.680 $\pm$ 0.016 & 5712 $\pm$ 3 & 4.395 $\pm$ 0.010 & 0.104 $\pm$ 0.003 & 0.160 $\pm$ 0.002 & -5.115 $\pm$ 0.017 &    5.9$^{+0.4}_{-0.4}$    & 56 &    6.0  \\ 
11915 & 8.615 $\pm$ 0.008 & 0.649 $\pm$ 0.003 & 5769 $\pm$ 4 & 4.480 $\pm$ 0.011 & -0.067 $\pm$ 0.004 & 0.187 $\pm$ 0.006 & -4.923 $\pm$ 0.028 &    3.6$^{+0.5}_{-0.7}$    & 54 &    6.0  \\ 
14501*   & 6.966 $\pm$ 0.007 & 0.645 $\pm$ 0.010 & 5738 $\pm$ 4 & 4.305 $\pm$ 0.012 & -0.153 $\pm$ 0.003 & 0.159 $\pm$ 0.001 & -5.108 $\pm$ 0.009 &    8.8$^{+0.3}_{-0.3}$    & 76 &    14.0  \\ 
14614 & 7.840 $\pm$ 0.010 & 0.620 $\pm$ 0.015 & 5803 $\pm$ 4 & 4.450 $\pm$ 0.013 & -0.109 $\pm$ 0.004 & 0.175 $\pm$ 0.003 & -4.977 $\pm$ 0.021 &    4.7$^{+0.4}_{-0.6}$    & 35 &    11.9  \\ 
15527 & 7.362 $\pm$ 0.004 & 0.650 $\pm$ 0.006 & 5779 $\pm$ 4 & 4.335 $\pm$ 0.011 & -0.064 $\pm$ 0.003 & 0.173 $\pm$ 0.003 & -5.000 $\pm$ 0.019 &    7.7$^{+0.4}_{-0.3}$    & 80 &    14.0  \\ 
18844*   & 6.739 $\pm$ 0.004 & 0.676 $\pm$ 0.002 & 5734 $\pm$ 3 & 4.365 $\pm$ 0.010 & 0.014 $\pm$ 0.003 & 0.158 $\pm$ 0.001 & -5.120 $\pm$ 0.007 &    7.0$^{+0.3}_{-0.4}$    & 16 &    1.2  \\ 
19911*   & 7.500 $\pm$ 0.010 & 0.661 $\pm$ 0.015 & 5761 $\pm$  -  &  --  &  --  & 0.264 $\pm$ 0.006 & -4.651 $\pm$ 0.016 &    --    & 7 &    1.2  \\ 
22263 & 5.497 $\pm$ 0.012 & 0.632 $\pm$ 0.006 & 5870 $\pm$ 7 & 4.535 $\pm$ 0.013 & 0.037 $\pm$ 0.006 & 0.276 $\pm$ 0.014 & -4.585 $\pm$ 0.034 &    0.8$^{+0.3}_{-0.4}$    & 137 &    6.0  \\ 
25670 & 8.275 $\pm$ 0.021 & 0.659 $\pm$ 0.015 & 5760 $\pm$ 3 & 4.420 $\pm$ 0.010 & 0.054 $\pm$ 0.003 & 0.168 $\pm$ 0.005 & -5.038 $\pm$ 0.033 &    5.1$^{+0.3}_{-0.7}$    & 59 &    6.0  \\ 
28066 & 6.592 $\pm$ 0.008 & 0.649 $\pm$ 0.010 & 5742 $\pm$ 4 & 4.300 $\pm$ 0.011 & -0.147 $\pm$ 0.003 & 0.158 $\pm$ 0.001 & -5.116 $\pm$ 0.007 &    8.8$^{+0.3}_{-0.3}$    & 84 &    6.0  \\ 
29432 & 6.861 $\pm$ 0.008 & 0.633 $\pm$ 0.007 & 5762 $\pm$ 3 & 4.450 $\pm$ 0.010 & -0.112 $\pm$ 0.003 & 0.177 $\pm$ 0.004 & -4.983 $\pm$ 0.024 &    5.2$^{+0.4}_{-0.4}$    & 95 &    6.8  \\ 
29525 & 6.442 $\pm$ 0.014 & 0.660 $\pm$ 0.005 & 5741 $\pm$ 9 & 4.520 $\pm$ 0.016 & -0.012 $\pm$ 0.007 & 0.335 $\pm$ 0.016 & -4.501 $\pm$ 0.031 &    0.8$^{+0.9}_{-0.3}$    & 5 &    12.7  \\ 
30037*   & 9.162 $\pm$ 0.015 & 0.682 $\pm$ 0.023 & 5666 $\pm$ 3 & 4.420 $\pm$ 0.011 & 0.007 $\pm$ 0.003 & 0.171 $\pm$ 0.006 & -5.053 $\pm$ 0.037 &    6.7$^{+0.5}_{-0.5}$    & 10 &    5.2  \\ 
30158 & 8.479 $\pm$ 0.013 & 0.746 $\pm$ 0.020 & 5678 $\pm$ 4 & 4.365 $\pm$ 0.011 & -0.004 $\pm$ 0.003 & 0.161 $\pm$ 0.002 & -5.118 $\pm$ 0.014 &    7.9$^{+0.3}_{-0.3}$    & 36 &    5.0  \\ 
30476 & 6.671 $\pm$ 0.004 & 0.675 $\pm$ 0.040 & 5709 $\pm$ 4 & 4.280 $\pm$ 0.011 & -0.033 $\pm$ 0.003 & 0.157 $\pm$ 0.001 & -5.141 $\pm$ 0.007 &    9.0$^{+0.3}_{-0.3}$    & 110 &    13.3  \\ 
30502 & 8.667 $\pm$ 0.015 & 0.664 $\pm$ 0.016 & 5731 $\pm$ 4 & 4.400 $\pm$ 0.013 & -0.057 $\pm$ 0.004 & 0.163 $\pm$ 0.002 & -5.085 $\pm$ 0.013 &    7.0$^{+0.4}_{-0.5}$    & 36 &    7.8  \\ 
33094 & 6.038 $\pm$ 0.003 & 0.712 $\pm$ 0.005 & 5629 $\pm$ 7 & 4.110 $\pm$ 0.016 & 0.023 $\pm$ 0.005 & 0.150 $\pm$ 0.001 & -5.229 $\pm$ 0.014 &    8.9$^{+0.3}_{-0.3}$    & 88 &    5.0  \\ 
34511 & 7.992 $\pm$ 0.010 & 0.630 $\pm$ 0.011 & 5812 $\pm$ 4 & 4.445 $\pm$ 0.012 & -0.091 $\pm$ 0.003 & 0.165 $\pm$ 0.001 & -5.032 $\pm$ 0.008 &    4.0$^{+0.5}_{-0.4}$    & 29 &    5.0  \\ 
36512 & 7.729 $\pm$ 0.011 & 0.656 $\pm$ 0.011 & 5744 $\pm$ 2 & 4.445 $\pm$ 0.008 & -0.126 $\pm$ 0.002 & 0.169 $\pm$ 0.002 & -5.038 $\pm$ 0.016 &    5.9$^{+0.4}_{-0.5}$    & 65 &    13.2  \\ 
36515 & 6.657 $\pm$ 0.004 & 0.641 $\pm$ 0.006 & 5855 $\pm$ 12 & 4.555 $\pm$ 0.023 & -0.029 $\pm$ 0.009 & 0.360 $\pm$ 0.019 & -4.420 $\pm$ 0.031 &    0.5$^{+0.3}_{-0.3}$    & 46 &    3.7  \\ 
38072 & 9.222 $\pm$ 0.002 & 0.648 $\pm$ 0.017 & 5860 $\pm$ 9 & 4.505 $\pm$ 0.018 & 0.085 $\pm$ 0.007 & 0.313 $\pm$ 0.013 & -4.504 $\pm$ 0.028 &    1.0$^{+0.8}_{-0.5}$    & 24 &    5.0  \\ 
40133 & 7.360 $\pm$ 0.012 & 0.660 $\pm$ 0.007 & 5745 $\pm$ 3 & 4.365 $\pm$ 0.009 & 0.116 $\pm$ 0.002 & 0.161 $\pm$ 0.002 & -5.095 $\pm$ 0.014 &    5.4$^{+0.3}_{-0.3}$    & 33 &    5.0  \\ 
41317 & 7.807 $\pm$ 0.004 & 0.668 $\pm$ 0.027 & 5706 $\pm$ 3 & 4.385 $\pm$ 0.010 & -0.081 $\pm$ 0.003 & 0.162 $\pm$ 0.001 & -5.103 $\pm$ 0.011 &    7.7$^{+0.3}_{-0.3}$    & 64 &    13.2  \\ 
42333 & 6.738 $\pm$ 0.008 & 0.655 $\pm$ 0.005 & 5846 $\pm$ 8 & 4.500 $\pm$ 0.016 & 0.132 $\pm$ 0.006 & 0.295 $\pm$ 0.018 & -4.547 $\pm$ 0.042 &    1.0$^{+0.7}_{-0.4}$    & 37 &    5.0  \\ 
43297 & 7.440 $\pm$ 0.008 & 0.689 $\pm$ 0.010 & 5705 $\pm$ 4 & 4.505 $\pm$ 0.009 & 0.082 $\pm$ 0.003 & 0.243 $\pm$ 0.022 & -4.730 $\pm$ 0.064 &    1.8$^{+0.5}_{-0.4}$    & 33 &    5.0  \\ 
44713 & 7.306 $\pm$ 0.006 & 0.668 $\pm$ 0.004 & 5759 $\pm$ 3 & 4.280 $\pm$ 0.010 & 0.063 $\pm$ 0.004 & 0.171 $\pm$ 0.005 & -5.019 $\pm$ 0.031 &    7.7$^{+0.3}_{-0.3}$    & 93 &    12.5  \\ 
44935 & 8.739 $\pm$ 0.014 & 0.645 $\pm$ 0.022 & 5771 $\pm$ 4 & 4.370 $\pm$ 0.012 & 0.038 $\pm$ 0.004 & 0.159 $\pm$ 0.003 & -5.095 $\pm$ 0.019 &    6.6$^{+0.3}_{-0.4}$    & 30 &    7.8  \\ 
44997 & 8.347 $\pm$ 0.023 & 0.659 $\pm$ 0.013 & 5728 $\pm$ 3 & 4.410 $\pm$ 0.011 & -0.012 $\pm$ 0.003 & 0.171 $\pm$ 0.006 & -5.033 $\pm$ 0.036 &    6.6$^{+0.4}_{-0.4}$    & 33 &    11.8  \\ 
49756 & 7.540 $\pm$ 0.008 & 0.647 $\pm$ 0.003 & 5789 $\pm$ 3 & 4.435 $\pm$ 0.009 & 0.023 $\pm$ 0.003 & 0.163 $\pm$ 0.002 & -5.058 $\pm$ 0.014 &    4.5$^{+0.3}_{-0.4}$    & 42 &    5.0  \\ 
54102*   & 8.653 $\pm$ 0.004 & 0.649 $\pm$ 0.018 & 5845 $\pm$ 6 & 4.510 $\pm$ 0.010 & 0.011 $\pm$ 0.005 & 0.228 $\pm$ 0.009 & -4.728 $\pm$ 0.029 &    0.7$^{+0.4}_{-0.4}$    & 26 &    11.0  \\ 
54287 & 7.223 $\pm$ 0.008 & 0.680 $\pm$ 0.002 & 5714 $\pm$ 4 & 4.340 $\pm$ 0.012 & 0.107 $\pm$ 0.004 & 0.157 $\pm$ 0.001 & -5.138 $\pm$ 0.009 &    6.5$^{+0.3}_{-0.4}$    & 74 &    13.2  \\ 
54582*   & 6.808 $\pm$ 0.008 & 0.613 $\pm$ 0.002 & 5883 $\pm$ 5 & 4.280 $\pm$ 0.014 & -0.096 $\pm$ 0.004 & 0.158 $\pm$ 0.001 & -5.056 $\pm$ 0.008 &    6.9$^{+0.3}_{-0.3}$    & 112 &    13.1  \\ 
62039*   & 7.817 $\pm$ 0.009 & 0.660 $\pm$ 0.004 & 5742 $\pm$ 3 & 4.340 $\pm$ 0.010 & 0.104 $\pm$ 0.003 & 0.157 $\pm$ 0.002 & -5.124 $\pm$ 0.012 &    6.2$^{+0.4}_{-0.3}$    & 41 &    5.0  \\ 
64150*   & 6.822 $\pm$ 0.061 & 0.676 $\pm$ 0.020 & 5747 $\pm$ 2 & 4.370 $\pm$ 0.008 & 0.049 $\pm$ 0.003 & 0.160 $\pm$ 0.001 & -5.100 $\pm$ 0.005 &    6.4$^{+0.3}_{-0.3}$    & 75 &    5.0  \\ 
64673 & 8.336 $\pm$ 0.010 & 0.640 $\pm$ 0.012 & 5912 $\pm$ 5 & 4.290 $\pm$ 0.014 & -0.017 $\pm$ 0.004 & 0.163 $\pm$ 0.003 & -5.007 $\pm$ 0.018 &    6.0$^{+0.4}_{-0.4}$    & 39 &    5.0  \\ 
64713 & 9.260 $\pm$ 0.022 & 0.649 $\pm$ 0.029 & 5788 $\pm$ 4 & 4.435 $\pm$ 0.013 & -0.043 $\pm$ 0.003 & 0.165 $\pm$ 0.004 & -5.046 $\pm$ 0.027 &    5.3$^{+0.5}_{-0.6}$    & 26 &    7.8  \\ 
65708*   & 7.426 $\pm$ 0.008 & 0.647 $\pm$ 0.011 & 5746 $\pm$ 5 & 4.220 $\pm$ 0.012 & -0.063 $\pm$ 0.005 & 0.155 $\pm$ 0.001 & -5.137 $\pm$ 0.005 &    9.0$^{+0.3}_{-0.3}$    & 10 &    1.1  \\ 
67620*   & 6.430 $\pm$  -  & 0.701 $\pm$ 0.050 & 5660 $\pm$  -  &  --  &  --  & 0.220 $\pm$ 0.006 & -4.823 $\pm$ 0.023 &    7.7$^{+0.6}_{-0.9}$    & 24 &    1.1  \\ 
68468 & 9.366 $\pm$ 0.025 & 0.654 $\pm$ 0.031 & 5845 $\pm$ 5 & 4.330 $\pm$ 0.013 & 0.071 $\pm$ 0.004 & 0.154 $\pm$ 0.003 & -5.103 $\pm$ 0.021 &    5.5$^{+0.3}_{-0.4}$    & 41 &    5.0  \\ 
69645 & 9.416 $\pm$ 0.014 & 0.665 $\pm$ 0.004 & 5751 $\pm$ 3 & 4.435 $\pm$ 0.010 & -0.026 $\pm$ 0.004 & 0.166 $\pm$ 0.003 & -5.052 $\pm$ 0.022 &    5.7$^{+0.3}_{-0.9}$    & 23 &    5.0  \\ 
72043*   & 7.511 $\pm$ 0.008 & 0.636 $\pm$ 0.010 & 5845 $\pm$ 4 & 4.340 $\pm$ 0.011 & -0.026 $\pm$ 0.003 & 0.178 $\pm$ 0.004 & -4.942 $\pm$ 0.020 &    6.2$^{+0.4}_{-0.3}$    & 31 &    5.0  \\ 
73241*   & 6.344 $\pm$ 0.008 & 0.71 $\pm$ 0.002 & 5661 $\pm$ 5 & 4.215 $\pm$ 0.014 & 0.092 $\pm$ 0.005 & 0.175 $\pm$ 0.006 & -5.032 $\pm$ 0.037 &    8.9$^{+0.3}_{-0.3}$    & 43 &    1.1  \\ 
73815 & 8.174 $\pm$ 0.003 & 0.663 $\pm$ 0.005 & 5790 $\pm$ 3 & 4.325 $\pm$ 0.008 & 0.023 $\pm$ 0.003 & 0.160 $\pm$ 0.002 & -5.080 $\pm$ 0.017 &    7.2$^{+0.3}_{-0.3}$    & 38 &    11.8  \\ 
74389 & 7.773 $\pm$ 0.014 & 0.636 $\pm$ 0.012 & 5845 $\pm$ 3 & 4.440 $\pm$ 0.011 & 0.083 $\pm$ 0.003 & 0.184 $\pm$ 0.005 & -4.909 $\pm$ 0.028 &    3.9$^{+0.3}_{-0.6}$    & 31 &    2.9  \\ 
74432 & 6.644 $\pm$ 0.008 & 0.682 $\pm$ -- & 5679 $\pm$ 5 & 4.170 $\pm$ 0.013 & 0.048 $\pm$ 0.005 & 0.150 $\pm$ 0.01 & -5.205 $\pm$ 0.059 &    8.6$^{+0.3}_{-0.3}$    & 59 &    4.8  \\ 
76114 & 7.217 $\pm$ 0.008 & 0.656 $\pm$ 0.007 & 5740 $\pm$ 3 & 4.410 $\pm$ 0.010 & -0.024 $\pm$ 0.003 & 0.162 $\pm$ 0.001 & -5.084 $\pm$ 0.008 &    6.6$^{+0.3}_{-0.3}$    & 33 &    4.8  \\ 
77052 & 5.868 $\pm$ 0.011 & 0.686 $\pm$ 0.002 & 5687 $\pm$ 3 & 4.450 $\pm$ 0.012 & 0.051 $\pm$ 0.003 & 0.211 $\pm$ 0.016 & -4.848 $\pm$ 0.060 &    4.5$^{+1.1}_{-0.4}$    & 152 &    11.8  \\ 
77883 & 8.755 $\pm$ 0.020 & 0.687 $\pm$ 0.024 & 5699 $\pm$ 3 & 4.375 $\pm$ 0.011 & 0.017 $\pm$ 0.003 & 0.164 $\pm$ 0.003 & -5.090 $\pm$ 0.020 &    7.6$^{+0.3}_{-0.4}$    & 32 &    7.8  \\ 
79578*   & 6.533 $\pm$ 0.033 & 0.647 $\pm$ 0.006 & 5810 $\pm$ 3 & 4.465 $\pm$ 0.010 & 0.048 $\pm$ 0.003 & 0.202 $\pm$ 0.008 & -4.839 $\pm$ 0.035 &    2.4$^{+0.6}_{-0.4}$    & 48 &    4.8  \\ 
79672 & 5.510 $\pm$ 0.009 & 0.650 $\pm$ 0.009 & 5808 $\pm$ 3 & 4.440 $\pm$ 0.009 & 0.041 $\pm$ 0.003 & 0.170 $\pm$ 0.004 & -5.005 $\pm$ 0.026 &    4.2$^{+0.3}_{-0.5}$    & 3775 &    13.2  \\ 
79715 & 8.357 $\pm$ 0.014 & 0.653 $\pm$ 0.019 & 5816 $\pm$ 4 & 4.380 $\pm$ 0.011 & -0.037 $\pm$ 0.004 & 0.161 $\pm$ 0.002 & -5.059 $\pm$ 0.015 &    6.2$^{+0.3}_{-0.4}$    & 34 &    4.8  \\ 
81746*   & 7.024 $\pm$ 0.008 & 0.653 $\pm$ -- & 5715 $\pm$ 3 & 4.370 $\pm$ 0.010 & -0.091 $\pm$ 0.003 & 0.160 $\pm$ 0.001 & -5.108 $\pm$ 0.008 &    8.1$^{+0.3}_{-0.3}$    & 14 &    4.8  \\ 
83276*   & 7.107 $\pm$ 0.004 &    -- & 5886 $\pm$ 6 & 4.240 $\pm$ 0.015 & -0.093 $\pm$ 0.005 & 0.155 $\pm$ 0.001 & -5.078 $\pm$ 0.009 &    7.4$^{+0.3}_{-0.3}$    & 6 &    0.3  \\ 
85042 & 6.287 $\pm$ 0.004 & 0.679 $\pm$ 0.001 & 5685 $\pm$ 3 & 4.350 $\pm$ 0.010 & 0.030 $\pm$ 0.003 & 0.159 $\pm$ 0.001 & -5.130 $\pm$ 0.011 &    7.8$^{+0.3}_{-0.3}$    & 188 &    10.9  \\ 
87769*   & 8.435 $\pm$ 0.010 & 0.685 $\pm$ 0.014 & 5828 $\pm$ 3 & 4.40 $\pm$ 0.010 & 0.072 $\pm$ 0.004 & 0.167 $\pm$ 0.004 & -5.016 $\pm$ 0.024 &    5.0$^{+0.4}_{-1.0}$    & 27 &    4.2  \\ 
89650 & 8.944 $\pm$ 0.001 & 0.643 $\pm$ 0.022 & 5851 $\pm$ 3 & 4.415 $\pm$ 0.011 & -0.015 $\pm$ 0.003 & 0.159 $\pm$ 0.003 & -5.058 $\pm$ 0.022 &    4.3$^{+0.7}_{-0.3}$    & 24 &    8.4  \\ 
95962 & 7.265 $\pm$ 0.008 & 0.643 $\pm$ 0.015 & 5805 $\pm$ 3 & 4.380 $\pm$ 0.009 & 0.029 $\pm$ 0.003 & 0.162 $\pm$ 0.002 & -5.056 $\pm$ 0.015 &    6.0$^{+0.4}_{-0.3}$    & 78 &    12.2  \\ 
96160 & 8.685 $\pm$ 0.013 & 0.653 $\pm$ 0.016 & 5798 $\pm$ 4 & 4.480 $\pm$ 0.012 & -0.036 $\pm$ 0.003 & 0.190 $\pm$ 0.004 & -4.899 $\pm$ 0.019 &    2.6$^{+0.4}_{-0.5}$    & 40 &    6.0  \\ 
101905 & 7.328 $\pm$ 0.008 & 0.626 $\pm$ 0.002 & 5906 $\pm$ 5 & 4.500 $\pm$ 0.011 & 0.088 $\pm$ 0.004 & 0.211 $\pm$ 0.015 & -4.769 $\pm$ 0.057 &    1.2$^{+0.3}_{-0.3}$    & 53 &    6.0  \\ 
102040 & 6.425 $\pm$ 0.004 & 0.613 $\pm$ 0.009 & 5853 $\pm$ 4 & 4.480 $\pm$ 0.012 & -0.080 $\pm$ 0.003 & 0.176 $\pm$ 0.004 & -4.950 $\pm$ 0.025 &    2.4$^{+0.4}_{-0.4}$    & 97 &    5.5  \\ 
102152 & 9.208 $\pm$ 0.015 & 0.669 $\pm$ 0.030 & 5718 $\pm$ 4 & 4.325 $\pm$ 0.011 & -0.016 $\pm$ 0.003 & 0.159 $\pm$ 0.002 & -5.118 $\pm$ 0.018 &    8.6$^{+0.3}_{-0.4}$    & 50 &    8.4  \\ 
103983*   & 8.390 $\pm$  -  &    -- & 5755 $\pm$  -  &  --  &  --  & 0.190 $\pm$ 0.007 & -4.917 $\pm$ 0.039 &    --    & 5 &    3.8  \\ 
104045 & 8.336 $\pm$ 0.014 & 0.639 $\pm$ 0.004 & 5826 $\pm$ 3 & 4.410 $\pm$ 0.010 & 0.051 $\pm$ 0.003 & 0.160 $\pm$ 0.003 & -5.066 $\pm$ 0.018 &    4.1$^{+0.9}_{-0.3}$    & 35 &    6.0  \\ 
105184 & 6.752 $\pm$ 0.004 & 0.640 $\pm$ 0.001 & 5843 $\pm$ 6 & 4.510 $\pm$ 0.011 & 0.003 $\pm$ 0.004 & 0.237 $\pm$ 0.011 & -4.702 $\pm$ 0.035 &    0.6$^{+0.5}_{-0.3}$    & 103 &    6.0  \\ 
108158 & 7.428 $\pm$ 0.008 & 0.696 $\pm$ 0.005 & 5675 $\pm$ 4 & 4.285 $\pm$ 0.011 & 0.055 $\pm$ 0.003 & 0.164 $\pm$ 0.001 & -5.102 $\pm$ 0.009 &    8.1$^{+0.3}_{-0.3}$    & 12 &    13.0  \\ 
108468 & 7.157 $\pm$ 0.008 & 0.625 $\pm$ 0.009 & 5841 $\pm$ 4 & 4.350 $\pm$ 0.011 & -0.096 $\pm$ 0.004 & 0.168 $\pm$ 0.003 & -5.002 $\pm$ 0.017 &    7.0$^{+0.3}_{-0.3}$    & 71 &    13.3  \\ 
109821 & 6.223 $\pm$ 0.009 & 0.652 $\pm$ 0.006 & 5747 $\pm$ 4 & 4.310 $\pm$ 0.011 & -0.108 $\pm$ 0.004 & 0.160 $\pm$ 0.002 & -5.101 $\pm$ 0.012 &    8.9$^{+0.3}_{-0.3}$    & 97 &    14.0  \\ 
114328 & 8.723 $\pm$ 0.014 & 0.676 $\pm$ 0.018 & 5775 $\pm$ 4 & 4.360 $\pm$ 0.012 & -0.017 $\pm$ 0.004 & 0.159 $\pm$ 0.003 & -5.093 $\pm$ 0.019 &    6.8$^{+0.4}_{-0.3}$    & 18 &    3.3  \\ 
114615 & 9.567 $\pm$ 0.035 & 0.656 $\pm$ 0.038 & 5819 $\pm$ 5 & 4.510 $\pm$ 0.009 & -0.063 $\pm$ 0.004 & 0.204 $\pm$ 0.006 & -4.829 $\pm$ 0.026 &    0.5$^{+1.2}_{-0.3}$    & 31 &    4.8  \\ 
115577 & 7.584 $\pm$ 0.008 & 0.692 $\pm$ 0.001 & 5694 $\pm$ 4 & 4.260 $\pm$ 0.010 & 0.013 $\pm$ 0.003 & 0.154 $\pm$ 0.001 & -5.165 $\pm$ 0.012 &    8.8$^{+0.3}_{-0.3}$    & 131 &    14.0  \\ 
116906*   & 7.682 $\pm$ 0.009 & 0.648 $\pm$ 0.002 & 5790 $\pm$ 3 & 4.370 $\pm$ 0.009 & -0.005 $\pm$ 0.003 & 0.161 $\pm$ 0.002 & -5.075 $\pm$ 0.011 &    6.7$^{+0.3}_{-0.3}$    & 26 &    9.4  \\ 
117367 & 7.676 $\pm$ 0.009 & 0.622 $\pm$ 0.007 & 5867 $\pm$ 3 & 4.350 $\pm$ 0.010 & 0.024 $\pm$ 0.003 & 0.159 $\pm$ 0.001 & -5.056 $\pm$ 0.009 &    5.7$^{+0.3}_{-0.3}$    & 43 &    6.0  \\ 
118115 & 7.889 $\pm$ 0.013 & 0.633 $\pm$ 0.002 & 5798 $\pm$ 4 & 4.275 $\pm$ 0.011 & -0.036 $\pm$ 0.003 & 0.157 $\pm$ 0.002 & -5.099 $\pm$ 0.012 &    8.0$^{+0.3}_{-0.3}$    & 47 &    6.0  \\

\hline
\end{longtable}}
\begin{longtable}{l}
* Spectroscopic binary \\
\end{longtable}
\end{longtab}
\setcounter{table}{2}

\begin{table*}[!htbp]
	
	\begin{center}	\caption{Chromospheric age errors as a function of time-span coverage. The age-activity outliers HIP15527 and HIP44713 were removed from this analysis.}
		
		\begin{tabular}{c c c c c c}
			\hline
			\hline 
			Minimum Time-span  & $\left\langle Age_{\rm HK} - Age_{\rm ISO}\right\rangle $ (Gyr) & $\left\langle (Age_{\rm HK} - Age_{\rm ISO})/Age_{\rm ISO}\right\rangle$  & Number & $Age_{\rm ISO}^{\rm min}$ & $Age_{\rm ISO}^{\rm max}$\\
			(yr) & (Gyr) & -- & of Stars & (Gyr) & (Gyr) \\
			\hline
			$\geq$ 5 & -0.4 $\pm$ 0.9 & 16 $\pm$ 6 \% & 33 & 2.6 & 9.0 \\ 
			$\geq$ 7 & -0.4 $\pm$ 1.0 & 15 $\pm$ 4 \% & 23 & 4.2 & 9.0 \\ 
			$\geq$ 10 & -0.5 $\pm$ 1.0 & 16 $\pm$ 4 \% & 14 & 4.2 & 9.0 \\ 
			$\geq$ 13 & +0.0 $\pm$ 0.9 & 13 $\pm$ 3 \% & 6 & 4.8 & 9.0 \\

			\hline
		\end{tabular}
	\end{center}
	\label{table:timespan}
\end{table*}


\end{document}